\documentclass[journal]{IEEEtran}
\usepackage{amsmath,amsfonts,amsthm}	
\usepackage[pdftex]{graphicx}														
\usepackage{url}
\usepackage{amsmath} 
\usepackage{amssymb} 
\usepackage{graphicx}
\usepackage{cite}
\usepackage[]{times}
\usepackage{epsfig}
\usepackage{subfigure}
\usepackage{bm}
\usepackage{wrapfig}
\usepackage{multirow}
\usepackage{color}
\usepackage{url}
\usepackage{verbatim}
\usepackage{cite}
\usepackage{algorithmic}
\usepackage{algorithm}

\newtheorem{theorem}{Theorem}

\newtheorem{lemma}{Lemma}
\newtheorem{definition}{Definition}
\newtheorem{corollary}{Corollary}
\ifCLASSINFOpdf
 
\else
  
\fi
\onecolumn
\begin{document}




\title{Check-hybrid GLDPC Codes: Systematic Elimination of Trapping Sets  and Guaranteed Error Correction Capability}
\author{Vida~Ravanmehr, 
Mehrdad Khatami, 
David Declercq, 
Bane Vasic

\thanks{This work was presented (in part) at the International Symposium on Information Theory (ISIT), Jun 28-July 4 2014, Honolulu, HI and Information Theory and Applications Workshop (ITA), Feb. 2014, San Diego, CA.}
\thanks{This work is funded by the Seagate Technology and in part by the NSF under grants CCF-0963726 and CCF-1314147.}
\thanks{V. Ravanmehr was with the Department of Electrical and Computer Engineering, University of Arizona. She is now  with the Coordinated Science Laboratory, University of Illinois, Urbana-Champaign, Urbana, IL, 61801 USA e-mail: vidarm@illinois.edu}   
\thanks{M. Khatami was the Department of Electrical and Computer Engineering, University of Arizona. He is now with the Marvell Semiconductor, Santa Clara,~CA, 95054,~USA email: khatami.mehrdad@gmail.com}
\thanks{D. Declercq is with ETIS,~ENSEA~/~University of Cergy-Pontoise~/~CNRS F-95000~Cergy-Pontoise,~France 
email: declercq@ensea.fr}
\thanks{B. Vasic is with the Department of Electrical and Computer Engineering, University of Arizona,~Tucson,~AZ, 85721,~USA email: vasic@ece.arizona.edu}} 


\maketitle
\begin{abstract}
In this paper, we propose a new approach to construct a class of check-hybrid generalized low-density parity-check (CH-GLDPC) codes  which are free of small trapping sets. The approach is based on converting some selected check nodes involving a trapping set into super checks corresponding to a 2-error correcting component code. Specifically, we follow two main purposes to construct the check-hybrid codes; first, based on the knowledge of the trapping sets of the global LDPC code,  single parity checks are replaced by super checks to disable the trapping sets. We show that by converting specified  single check nodes, denoted as critical checks, to super checks in a trapping set, the parallel bit flipping (PBF) decoder  corrects the errors on a trapping set and hence eliminates the trapping set. The second purpose is to minimize the rate loss caused by replacing the super checks  through finding the minimum number of such critical checks. We also present an algorithm to find critical checks in a trapping set of  column-weight 3  LDPC code and then provide  upper bounds on the minimum number of such critical checks  such that  the decoder  corrects all error patterns on elementary trapping sets. Moreover, we provide a fixed set for a  class of constructed check-hybrid codes. The guaranteed error correction capability of the CH-GLDPC codes is also studied. We show that a CH-GLDPC code in which each variable node is connected to 2 super checks corresponding to a 2-error correcting component code corrects up to 5 errors. The results are also  extended to column-weight 4 LDPC codes. Finally, we investigate the eliminating of trapping sets of a column-weight 3 LDPC code using the Gallager B decoding algorithm and generalize the results obtained for the PBF for the Gallager B decoding algorithm.

\begin{IEEEkeywords}  Check-hybrid GLDPC codes,  Critical set, Error correction capability, Gallager B decoding algorithm, Low-density parity-check (LDPC) codes, Parallel bit flipping (PBF) algorithm, Splitting number, Trapping set.  
\end{IEEEkeywords}

\end{abstract}
\IEEEpeerreviewmaketitle

\section{Introduction}

\IEEEPARstart{I}{t} has been shown that short low-rate codes with a good performance can be constructed from generalized low-density parity-check (GLDPC) codes with hybrid check nodes (e.g. \cite{Liva_Hamming}, \cite{Liva_QC}). 
Liva and Ryan \cite{Liva_Hamming} were first who defined  {\it{doping}}   to refer to substituting some single parity checks by super checks corresponding to a stronger linear block code and constructed check-hybrid GLDPC (CH-GLDPC) codes using Hamming codes as component codes. In another work by Liva {\it{et al.}} \cite{Liva_QC},  low-rate GLDPC codes are constructed by doping quasi-cyclic (QC)-LDPC codes  with Hamming codes. It was shown that the constructed codes have a remarkable performance both in the waterfall and the error-floor regions on the additive white Gaussian noise (AWGN) channel. 
 Paolini {\it{et al.}} \cite{Paolini_1}, \cite{Paolini_2} studied the GLDPC  and doubly-GLDPC codes with Hamming or  BCH codes as the component codes  and proposed a method for the asymptotic analysis of doubly-GLDPC codes on the binary erasure channel (BEC). They also considered CH-GLDPC codes and showed that the asymptotic threshold of hybrid GLDPC codes outperforms that of the LDPC codes. 
In another work \cite{Paolini_3}, Paolini {\it{et al.}} analyzed the asymptotic exponent of both the weight spectrum and the stopping set size spectrum for the CH-GLDPC codes and provided a simple formula for the asymptotic exponent of the weight distribution of the CH-GLDPC codes.  
 Two common features of the methods given in the previous work are: (i) replacing the super checks based on  degree distribution or density evolution of the resulting CH-GLDPC codes, and (ii) significant reduction of the rate of   CH-GLDPC codes compared to the original LDPC code. In this paper, we propose a method to construct CH-GLDPC codes; however, our approach  is different in that the super checks are chosen specifically to address the error floor issue and is based on the knowledge of failures of the global 
LDPC code on the BSC under  the parallel bit flipping (PBF) algorithm.  The PBF algorithm is a simple algorithm with low complexity and hence suitable for high-speed applications. This algorithm is also appropriate for the analysis of failures of iterative decoding algorithms of  LDPC codes, first identified by Richardson and denoted as ``trapping sets" \cite{Richardson}. While  trapping sets of the LDPC codes over the binary erasure channel (BEC) are well characterized as ``stopping sets", they are more complicated to define over the BSC and the AWGN channel. In \cite{Vasic},  the most harmful structures of column-weight three LDPC codes on the BSC using Gallager A/B and the PBF  algorithms have been identified. It was also shown that the trapping sets are short cycles or can be obtained as the union of short cycles in the Tanner graph. 
   One important aspect of this work is to provide a guidance in order to jointly design the Tanner graph of the proposed CH-GLDPC codes, and assign the location of the component codes with the objective of lowering the error floor. Our construction of  the CH-GLDPC codes is decomposed in two steps: we start with a classical LDPC code design (QC, protograph, etc.), and the knowledge of its small trapping sets, then,  instead of randomly choosing super checks, we place the super checks corresponding to a 2-error correcting component codes at those check nodes so that the PBF decoder can correct the errors on a trapping set. For an efficient check-hybrid code design, it is also desirable to find the minimum number of super checks such that the rate loss of the constructed check-hybrid codes be reduced.  In this paper, we study the minimum number of such critical super checks, denoted as the  {\it{splitting number}} and provide upper bounds on the splitting number for some dominant trapping sets. The LDPC codes that are used in this paper are column-weight three and column-weight four LDPC codes. We first focus on trapping sets of column-weight three LDPC codes and provide an algorithm to find critical checks in a trapping set and also provide upper bounds on the splitting number of trapping sets. Furthermore, we study the error correction capability of two classes of CH-GLDPC codes using a column-weight three LDPC code as the global code and show that a CH-GLDPC code in which each variable node is connected to 2 super checks is able to correct up to 5 errors. The results obtained for the critical checks, splitting number and error correction capability of CH-GLDPC codes  with column-weight three LDPC codes as the global code and the PBF decoding algorithm are generalized when the Gallager B decoding algorithm is used. 
 
 The rest of the paper is organized as follows. In Section \ref{Pre}, we provide the notations and definitions that are used throughout the paper. In section \ref{SuperChecks}, we characterize the effect of super checks in terms of trapping sets elimination. In section \ref{results}, we present our main results on  CH-GLDPC codes free of small trapping sets. In section \ref{GEC}, we give the guaranteed error correction capability of the constructed CH-GLDPC codes. In section \ref{Discussion}, we extend some of our results for column-weight  four global LDPC codes and also Gallager B decoder. 
 Section \ref{conclusion} concludes the paper.
\section{Preliminaries}
\label{Pre}
In this section, we first  establish the notations and then give a brief summary on the definitions and concepts of LDPC and GLDPC codes. We also define trapping sets and fixed sets for the iterative  decoding algorithms. 
\subsection{Graph Theory Notations }
Let $G(U,E)$ be an undirected simple graph with the set of vertices $U$ and the set of edges $E$. An edge $e$ is an unordered pair $(u_1,u_2)$. The edge $e=(u_1,u_2)$ is said to be incident on $u_1$ and $u_2$ and the two vertices $u_1$ and $u_2$ are said to be adjacent (neighbors). The set of neighbors of the vertex $u$ is denoted by ${\cal{N}}(u)$. The degree of each vertex $d(u)$ is defined as the number of vertices in its neighborhood. The length of the shortest cycle is called the girth of the graph and is denoted by $g$. A bipartite graph $G(V \cup C,E)$ is graph with two disjoint sets of vertices; variable nodes $V$ and check nodes $C$. An edge $e$ is incident on a variable node $v\in V$ and a check node $c \in C$. A bipartite graph is called $(\gamma,\rho)$-regular if the degree of each variable node is $\gamma$ and the degree of each check node is $\rho$. The girth of a bipartite graph  is even. 
The parity check matrix $H$ of a linear code $C$ can be represented with a bipartite graph called the Tanner graph. Each column in the parity check matrix is shown by a variable node and each row is denoted by a check node in the Tanner graph. A variable node $v_j$ and a check node $c_i$ are adjacent if and only if $H_{i,j}=1$.   A vector ${\bf{v}}=(v_1,v_2,...,v_n)$ is a codeword  if and only if  $H{\bf{v}}^{T}={\bf{0}} ~~({\rm{mod ~2}})$.  A linear code is called $(\gamma,\rho)$-regular if its parity check matrix is $(\gamma,\rho)$-regular. This code has rate $r \geq 1-\frac{\gamma}{\rho}$ \cite{Gallager}.
\subsection{LDPC codes, GLDPC and CH-GLDPC codes }
LDPC codes were first introduced by Gallager  in his landmark work \cite{Gallager} where he proposed different methods for constructing parity check matrices of LDPC codes and provided different hard decision algorithms for decoding of LDPC codes.  LDPC codes are usually defined by their Tanner graphs. A $(\gamma,\rho,g)$ LDPC code is a  $(\gamma,\rho)$-regular code of girth $g$.

 GLDPC codes were introduced by Tanner in \cite{Tanner_GLDPC} where he proposed a method to construct longer error-correcting codes from shorter error-correcting codes. In GLDPC codes, each super check node is satisfied if its neighboring variable nodes form a codeword of a linear code called {\it{component code}}. That is if  $c_i$ is a single parity check node in the Tanner graph of the global code and  $\{v_{i_1},v_{i_2},...,v_{i_n}\}$ with values $\{x_1,x_2,...,x_n\}$ are the neighbors of $c_i$, then in the GLDPC code, the super check corresponding to $c_i$ is satisfied if $(x_1,x_2,...,x_n)$ be a codeword of the component code. 
The parity check matrix of GLDPC codes is constructed using the parity check matrix of the longer code also known as the global code and the parity check matrix of the component code.  To construct the parity check matrix of the GLDPC code, it is enough to replace each one in each row of the parity check matrix of the global code by one column of the parity check matrix of the component code. Each zero in each row will be replaced by a zero-column in the  parity check matrix.

A CH-GLDPC code has two types of check nodes: single parity checks and super checks corresponding to a component code. As in GLDPC codes, a super check node is satisfied when its neighboring variable nodes be codeword of the component code, while the single parity check is satisfied when the modulo-2 sum of its neighboring variable nodes is zero. 
The component codes in GLDPC and CH-GLDPC codes can be chosen arbitrarily and possibly from different block codes. However, in this paper, GLDPC  and CH-GLDPC codes are constructed from the same component code and the global codes are chosen from the family of $(\gamma,\rho)$-regular codes. 
\subsection{Decoding Algorithms and Trapping Sets }
The decoding algorithms for decoding LDPC codes include a class of iterative algorithms such as bit flipping algorithms (parallel and serial) and messages passing algorithms like Gallager A/B and belief propagation decoding algorithms.

The notion of ``trapping sets" was first introduced by Richardson \cite{Richardson} as the structures in the Tanner graph of LDPC codes responsible for failures of decoders. Before we characterize the trapping sets of bit flipping decoding algorithm, we provide definitions and assumptions. 
In this paper, we consider transmission over the BSC. We also consider that the all-zero codeword is sent. Under this assumption, a variable node is said to be correct if its received value is 0; otherwise it is called corrupt. The support of a vector ${\bf{x}}=(x_1,x_2,...,x_n)$ denoted by ${\rm{supp}}({\bf{x}})$ is the set $\{i ~|~x_i \neq 0\}$. The decoder runs until the maximum number of iterations $M$ is reached or a codeword is found. Let ${\bf{y}}=(y_1,y_2,...,y_n)$ be a received vector after transmitting the all-zero codeword and let ${\bf{y}^{(l)}}=(y_1^{(l)},y_2^{(l)},...,y_n^{(l)})$ be the output of the decoder after the $l$-th iteration. A variable node $v$ is said to be eventually correct if there exists an integer $L>0$ such that for all $l\geq L$, $v \notin {\rm{supp}}({\bf{x}}^l)$. The decoder fails on decoding ${\bf{y}}$ if there does not exist $l \leq M$ such that $|{\rm{supp({\bf{x}})}}| = 0$. For the received word $\bf{y}$, the set of variable nodes which are not eventually correct is called a trapping set and is denoted by ${T}(\bf{y})$. If ${T}(\bf{y}) \neq \emptyset$,  then ${T}({\bf{y}})$ is called an $(a,b)$ trapping set and is denoted by ${\cal{T}}(a,b)$ if the number of variable nodes in ${T}({\bf{y}})$ equals $a$ and the number of odd degree check nodes in the subgraph induced by ${T}({\bf{y}})$ is $b$. For the trapping set $T({\bf{y}})$, ${\rm{supp({\bf{y}})}}$ is an induced set. ${\cal{T}}(a,b)$ is called an {\it{elementary trapping set}} if the degree of each check node in the subgraph induced by the set of variable nodes is one or two and there $b$ check nodes of degree one.

Chilappagari {\it{et al.}} \cite{Shashi_Error_Floor} introduced the notion of ``critical number" as the minimum number of variable nodes on a trapping set that need to be initially in error such that the decoder fails. It was shown that the harmfulness of a trapping set depends on its critical number; the smaller the critical number, the more harmful a trapping set. In this paper, we say that a trapping set is {\it{harmful}} if the decoder fails to decode at least one error pattern on the trapping set; Otherwise, it is called {\it{harmless}}.
While trapping sets can have different induced sets, a class of trapping sets called {\it{fixed sets}} have the fixed induced set. A fixed set $F$ is the set of  variable nodes that are corrupt at the beginning and at the end of decoding iterations, while the  variable nodes that are initially correct  remain correct after decoding. A vector $\bf{y}$ is called a fixed point if ${\rm{supp(y)}}=F$. From definition of the fixed set and trapping set, it is clear that a fixed set is always a trapping set while a trapping set is not necessarily a fixed set. Fixed sets of an LDPC code with the column-weight $\gamma$ are the set of variable nodes ${\cal{I}}$ such that every variable node in ${\cal{I}}$ is connected to at least $\left\lceil \gamma /2 \right\rceil$ of check nodes of even-degree and no $\left\lfloor \gamma /2 \right\rfloor$ check nodes of odd-degree share a variable node outside ${\cal{I}}$ \cite{Vasic}.
Chilappagari {\it{et al.}} defined   fixed sets for the PBF algorithm of GLDPC codes as follows:

{\bf{Fact 1}}:(\cite{shashi_GLDPC} Theorem 6) Let ${\cal{C}}$ be a GLDPC code with $(\gamma,\rho)$-regular global code and a $t$-error correcting component code. Let ${\cal{I}}$ be a  subset of variable nodes with the following properties: (a) The degree of each check node in ${\cal{I}}$ is either 1 or $t+1$; (b) Each variable node in ${\cal{I}}$ is connected to $\left\lceil \gamma /2 \right\rceil$ checks of degree $t+1$ and  $\left\lfloor \gamma /2 \right\rfloor$ check nodes of degree 1; and (c) No $\left\lfloor \gamma /2 \right\rfloor +1$ checks of degree $t+1$ share a variable node outside   ${\cal{I}}$. Then, ${\cal{I}}$ is a fixed set.
 \section{Effect of Super checks on trapping sets}
 \label{SuperChecks}
 
 Let us start by some observations on the effect of replacing single parity checks by super checks. In fact, we show how trapping sets responsible for the failure of the PBF are not harmful anymore when  some selected single checks are replaced by super checks \cite{RDV_14_ITA}, \cite{RDV_14_ISIT}.  We first describe the PBF algorithm for the CH-GLDPC codes and use it throughout the paper for our analysis. We mention that the decoding algorithm at each super check is the bounded distance decoding (BDD). The BDD is capable of correcting $t$ errors when the minimum distance of the code is at least $2t+1$.
 
 %
  \begin{algorithm}
\caption{The PBF algorithm for decoding CH-GLDPC codes \cite{RDV_14_ITA}, \cite{RDV_14_ISIT}.}
\label{Alg1}
\begin{algorithmic}
\STATE {\bf{In each iteration:}}
\begin{itemize}
\item  Variable nodes send their current estimates to the neighboring single parity check and super check nodes. 
\end{itemize}
\STATE {\bf{~~Updating rule at check nodes:}}
 \begin{itemize}
 \item Each super check node performs the BDD on the incoming messages. If a codeword is found, then the check node sends  flip messages to all variable nodes which differ from  the codeword. If not, then the check node does not send any flip messages.
 \end{itemize}
 \begin{itemize}
 \item At each single parity check, the modulo-2 sum of the incoming messages is calculated. If the sum is not zero, then the check node sends flip messages to the neighboring variable nodes. If the sum is zero, then the check node does not send any flip messages.
 \end{itemize}
\STATE {\bf{~~Updating rule at variable nodes:}}
\begin{itemize} \item A variable node flips if it receives more than $\gamma/2$ flip messages.
\end{itemize}
\end{algorithmic}
\end{algorithm} 

\begin{figure}[t]
       \begin{center}
        \subfigure[]{
         \centering\includegraphics[width=1.1in]{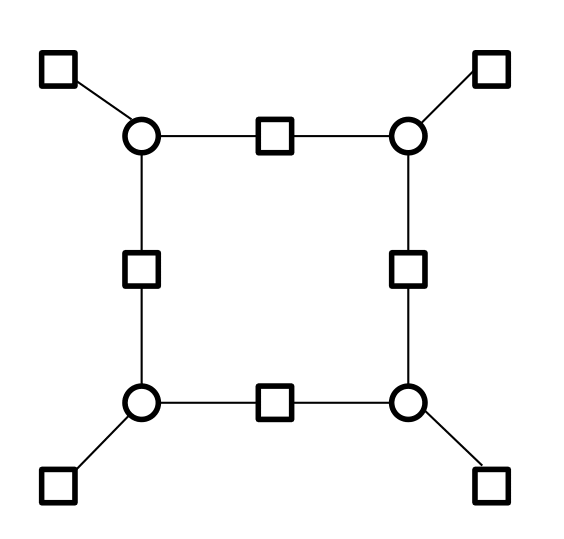}
         }
        \subfigure[]{
        \centering \includegraphics[width=1in]{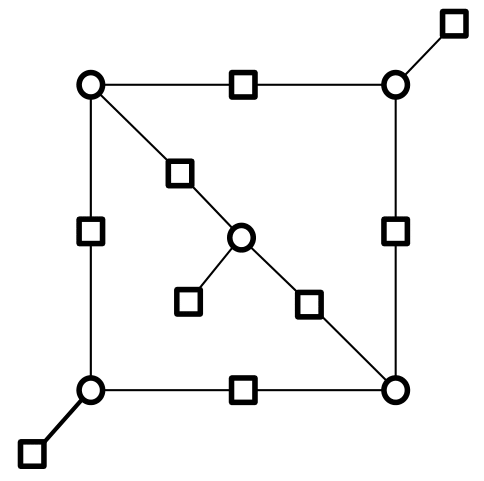}
              
   }
      \subfigure[]{
       \centering\includegraphics[width=1.8in]{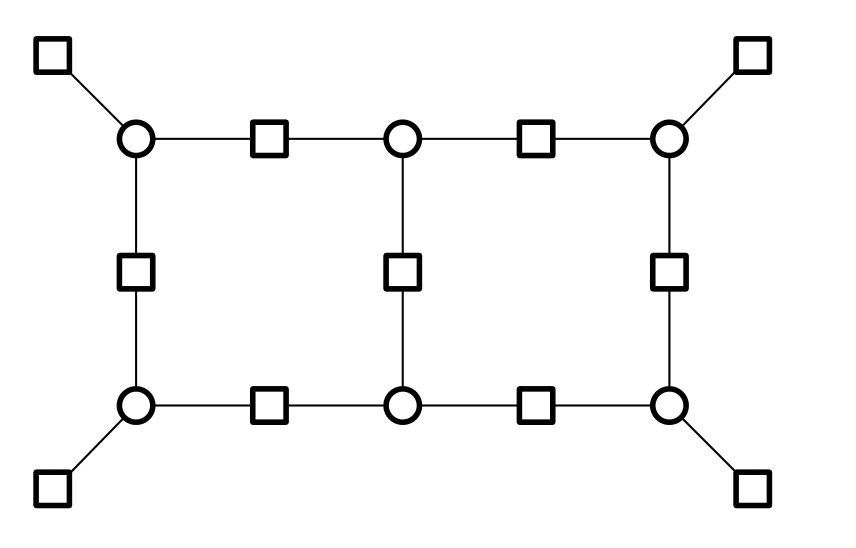}
         }

        \end{center}
        \caption{Tanner graph representation of trapping sets for column-weight three and girth $g=8$ LDPC codes; (a) the (4,4) trapping set, (b) the (5,3) trapping set, (c) a (6,4) trapping set.}
       \label{trapsets}
\end{figure}

\begin{figure}[t]
       \begin{center}
        \subfigure[]{
         \centering\includegraphics[width=1in]{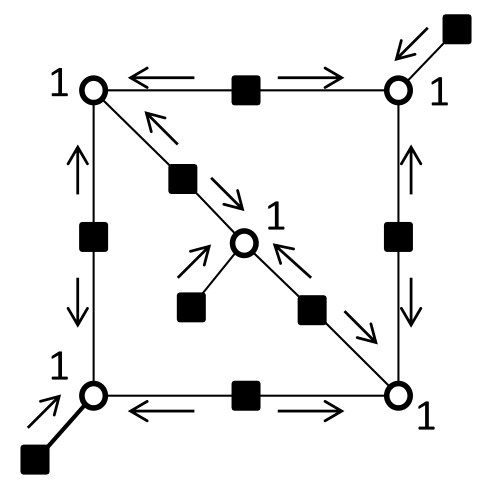}
         }
        \subfigure[]{
        \centering \includegraphics[width=1in]{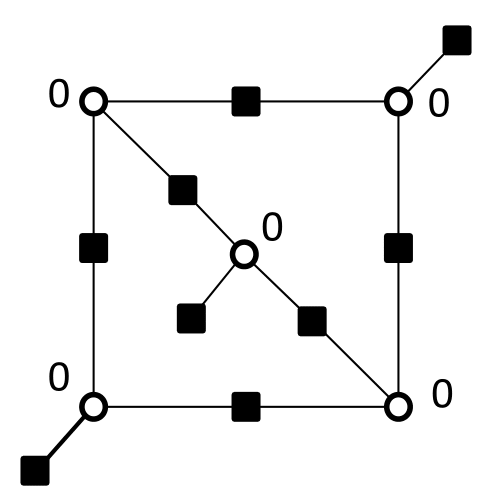}
              
   }

        \end{center}
        \caption{ The (5,3) trapping set is eliminated  when all single parity checks are replaced by super checks corresponding to a 2-error correcting component code. (a) flip messages from super checks to corrupt variable nodes in the first iteration of the PBF algorithm, (b) all variable nodes are corrected after the first iteration. }
        \label{AllSC}
\end{figure}

Let $\cal{C}$ be a $(3,\rho,8)$ LDPC code. 
Fig. \ref{trapsets} shows some small trapping sets of a column-weight three LDPC codes of girth $g=8$ namely the $(4,4)$ trapping set, the $(5,3)$ trapping set and a $(6,4)$ trapping set. In this paper, $\circ$ denotes a variable node and $\Box$ denotes a check node. 
It can be easily seen that if all single parity checks in the Tanner graph corresponding to the parity check matrix of ${\cal{C}}$ are replaced by super checks of a 2-error correcting component code, then the PBF decoding algorithm for GLDPC codes can correct all errors on the trapping sets. This result can be explained by the fact that in all elementary  trapping sets, the degree of each check node is at most two and since they are replaced by a 2-error correcting component code, the BDD at each super check can correct all errors. Fig. \ref{AllSC} shows how the PBF corrects all errors located on the (5,3) trapping set when all single checks are replaced by super checks. In this paper, a $\blacksquare$ denotes a super check and flip messages are shown with $\rightarrow$. However, as we show in the following, it is not necessary to replace all super checks in a trapping set for the decoder to correct the errors.  We show that it is possible to make the trapping set
harmless by replacing only some selected single checks by super checks. We say a trapping set is {\it{eliminated}} if by replacing super checks, the trapping set is not  harmful anymore.

Let consider the (5,3) trapping set. Fig. \ref{fig:ts(5,3)} shows how the PBF algorithm  corrects all errors located on the trapping set in which only two single parity checks of degree 2 are replaced by super checks.  
\begin{figure}[t]
        \begin{center}
        \subfigure[]{
         \centering\includegraphics[width=1.2in]{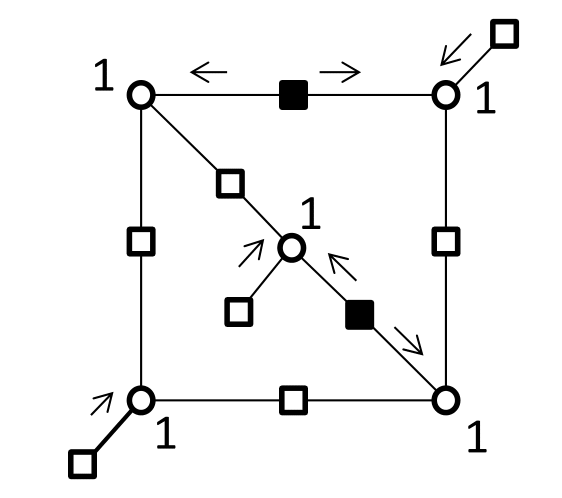}
         }
        \subfigure[]{
        \centering \includegraphics[width=1in]{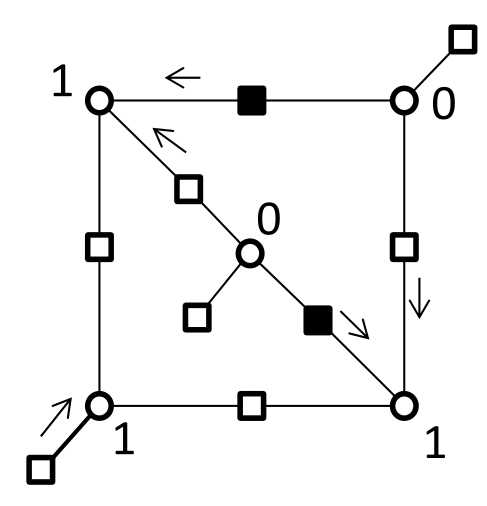}
              
   }
         \subfigure[]{
         \centering\includegraphics[width=1in]{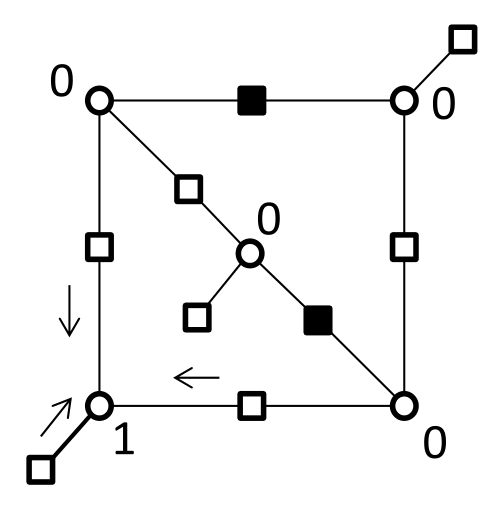}
         }
        \subfigure[]{
         \centering\includegraphics[width=1in]{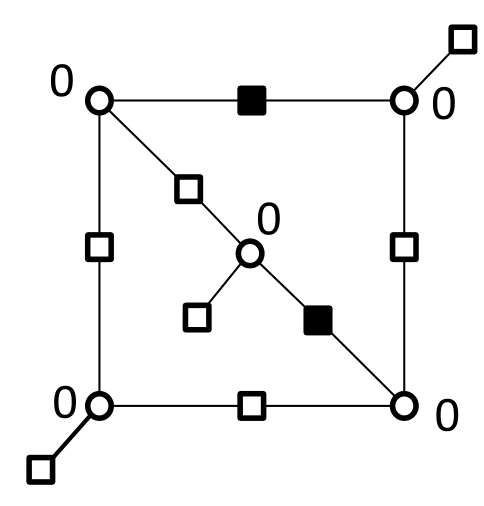}
         }
        \end{center}
        \caption{The (5,3) trapping set in a column-weight three code is eliminated if two super checks  corresponding to a 2-error correcting component code are replaced. Arrows show flip messages from check nodes to corrupt variable nodes in each iteration of the PBF algorithm: (a) flip messages from checks in the first iteration, (b) flip messages from checks to 3 variable nodes that are still in error, (c) flip messages from checks to the only one corrupt variable, (d) all variable nodes are corrected after the third iteration.}
        \label{fig:ts(5,3)}
\end{figure}

\begin{figure}[t]
        \begin{center}
        \subfigure[]{
         \centering\includegraphics[width=1.1in]{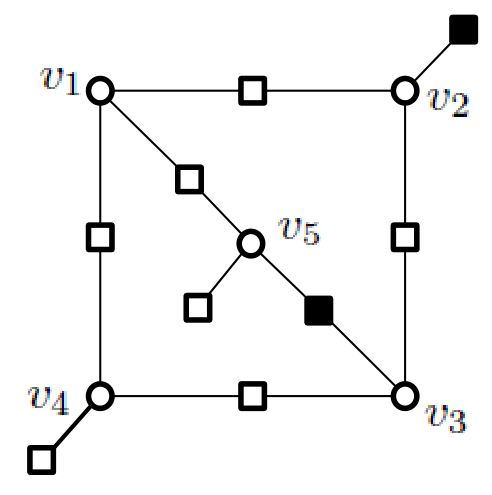}
         \label{exc1}
         }
        \subfigure[]{
         \centering \includegraphics[width=1.1in]{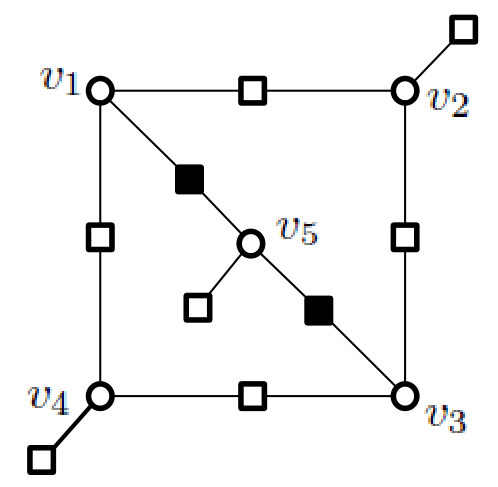}
         \label{exc2}     
   }
        \subfigure[]{
         \centering \includegraphics[width=1.1in]{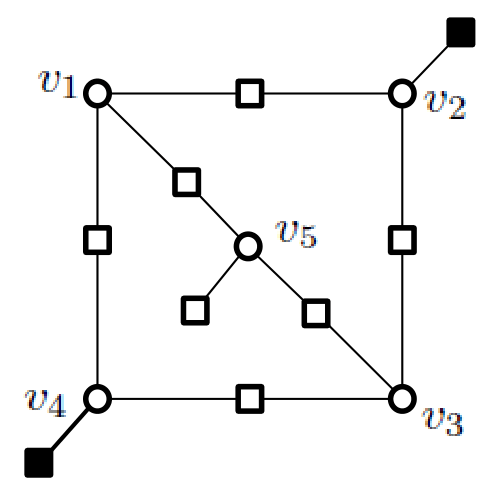}
         \label{exc3}     
   } 
        \end{center}
        \caption{Possible super-check replacements which are not helpful for the decoder to correct all errors on the (5,3) trapping set.}
        \label{exceptions}
\end{figure}

It should be noted that not all pairs of super checks in the (5,3) trapping set can be helpful for the decoder to correct the errors on the (5,3) trapping set. Fig. \ref{exceptions} shows three possible cases that by replacing the super checks  the trapping sets  remain harmful \cite{RDV_14_ITA}. In Fig. \ref{exc1} and \ref{exc2} only the variable node $v_5$ will be corrected, while in Fig. \ref{exc3} all variable nodes will remain incorrect. 

The above examples show that not only the number of super checks, but also the positions of super checks in a trapping set  are important for the decoder to successfully correct the errors.  Since the rate of the GLDPC codes decreases by replacing single parity checks by super checks, we are interested in replacing the minimum number of super checks such that the resulting Tanner graph will be free of small trapping sets.  In the next section, we first provide an algorithm to find a set of such critical checks in a trapping set and then   
we present upper bounds on the minimum number of super checks that need to be replaced in the parity check matrix  such that the resulting Tanner graph will be free of small trapping sets. 

\section{Critical sets and the splitting number}
\label{results}
In this section, we provide our main results on  CH-GLDPC codes in which the trapping sets responsible for the failure of the PBF algorithm have been eliminated. In this section, whenever not stated, the global LDPC code of the CH-GLDPC codes is a  $(3,\rho,8)$ LDPC code.
\subsection{ Critical sets and minimal size of critical sets }
As shown in Section \ref{SuperChecks}, a trapping set can be eliminated by judiciously replacing  check nodes in the original global code. A set of such checks is called a {\it{critical set}} and defined as follows \cite{RDV_14_ITA},\cite{RDV_14_ISIT}.
\begin{definition}
Let ${\cal{T}}(a,b)$ be an elementary trapping set. Let $C=\{c_1,c_2,...,c_k\}$ where $k\leq b$ be a set of check nodes of degree 2 in ${\cal{T}}$. A  set $S \subseteq C$ is called critical if by converting the single parity checks in $S$ to the super checks, the trapping set is eliminated. 
\end{definition} 
We note that a critical set is not unique  and there are many possible critical sets with different sizes in a trapping set.
\begin{definition}
\label{sp}
Let ${\cal{T}}(a,b)$ be an elementary trapping set. The minimum size of a critical set in ${\cal{T}}$  is denoted by $s_{(a,b)}({\cal{T}})$. 
\end{definition} 
As an example,$s_{(5,3)}({\cal{T}})=2$ as can be seen in Fig. \ref{fig:ts(5,3)}. 
In Algorithm \ref{Alg2}, we provide a method to find one of many possible critical sets in a trapping set.  The motivation behind finding a critical set using Algorithm \ref{Alg2} is based on the role of super checks in elementary trapping sets. When a single parity check of degree-2 is replaced by a super check, then the super check sends a flip message to a neighboring variable node if and only if the variable node is corrupt. Thus, each super check plays the role of 2 equivalent and isolated single parity checks, one for each of connected variable nodes. Breaking the cycles in a trapping set by splitting the super check into two single parity checks is the basis for finding a critical set in Algorithm \ref{Alg2}. 

\begin{figure}[t]
      
         \centering \includegraphics[width=2.4in]{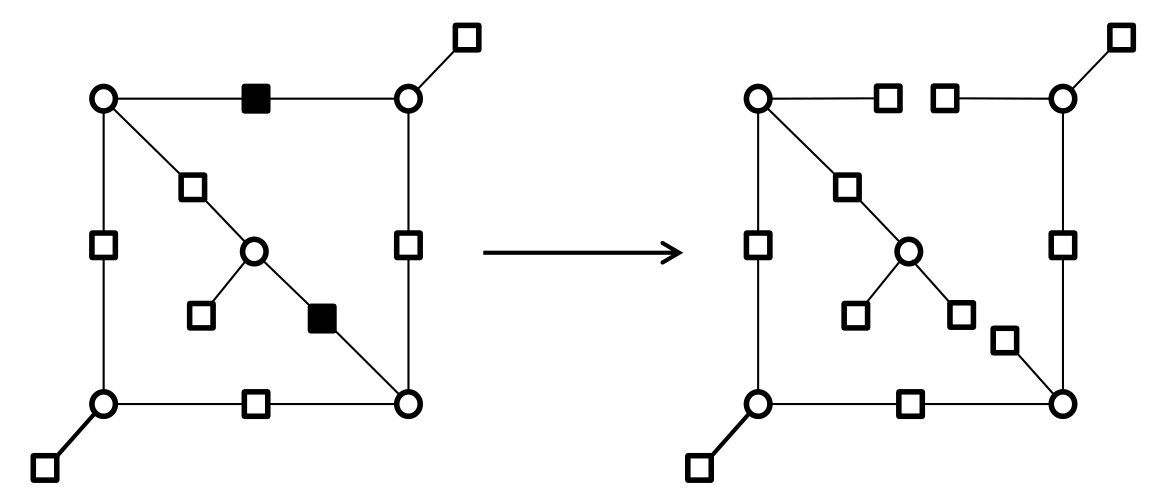}
      \caption{Super checks corresponding to a 2-error correcting component code can be considered as two single parity checks of degree-1. These replacements  break the cycles responsible for the failure of decoding.}
        \label{cycle breaking}
\end{figure}

 Fig. \ref{cycle breaking} shows an alternative  view of the effect of a super check  to eliminating a trapping set.

\begin{algorithm}
\caption{Finding a critical set in a trapping set ${\cal{T}}(a,b)$ \cite{RDV_14_ITA}, \cite{RDV_14_ISIT}.}
\label{Alg2}
\begin{algorithmic}
\STATE {\bf{initialization}:} Let ${\cal{T}}'={\cal{T}}$ be the $(a,b)$ trapping set.
\WHILE{Number of variable nodes in ${\cal{T'}}$ is greater than 0}
\IF{there exists a variable node $v$ in ${\cal{T'}}$ which is connected to exactly one degree-1 check node and two degree-2 checks }
\STATE Replace one of the check nodes of degree-2 connected to $v$ by a super check 
  corresponding to a 2-error correcting code. Split the super check into two single checks. Remove the variable node $v$ and all edges connected to it.
\ELSE
\STATE  Choose a variable node $v$ in ${\cal{T'}}$. Replace one check node of degree-2 connected to $v$ by a super check and split the super check node to 2 single parity checks.
\ENDIF
\WHILE{Number of variable nodes connected to at least two single parity checks of degree-1 is greater than 0} 
\STATE Remove variable nodes connected to at least two single parity checks of degree-1 and all edges connected to them.
\ENDWHILE
\ENDWHILE
\end{algorithmic}
\end{algorithm}

As we explained, the number of cycles in a trapping set plays a key role in finding the number of critical checks of a trapping set. This fact helps us to find the number of critical checks in some trapping sets without using Algorithm \ref{Alg2}. If a trapping set ${\cal{T'}}(a',b')$ has been obtained by adding some variable and check nodes to another trapping set ${\cal{T}}(a,b)$ such that the new variable and check nodes do not create a new cycle, then  $s_{(a',b')}({\cal{T'}})$ and $s_{(a,b)}({\cal{T}})$ are equal. To be more precise, we first provide the following definitions.

\begin{figure}[t]
        \begin{center}
        \subfigure[]{
         \centering\includegraphics[width=1.6in]{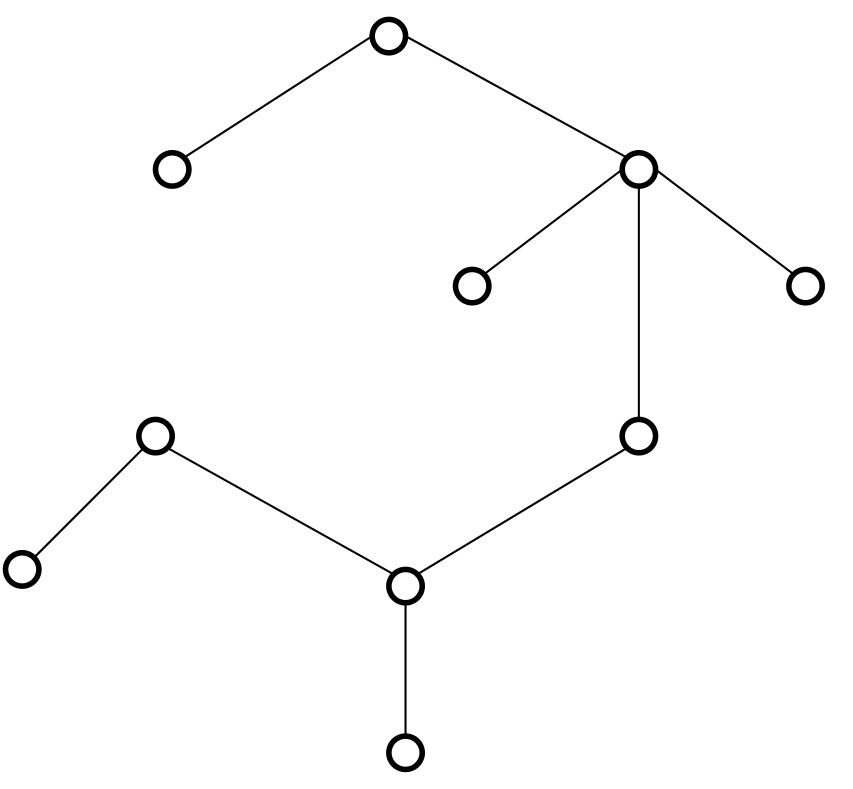}
         \label{simple}
         }
        \subfigure[]{
                \centering \includegraphics[width=1.6in]{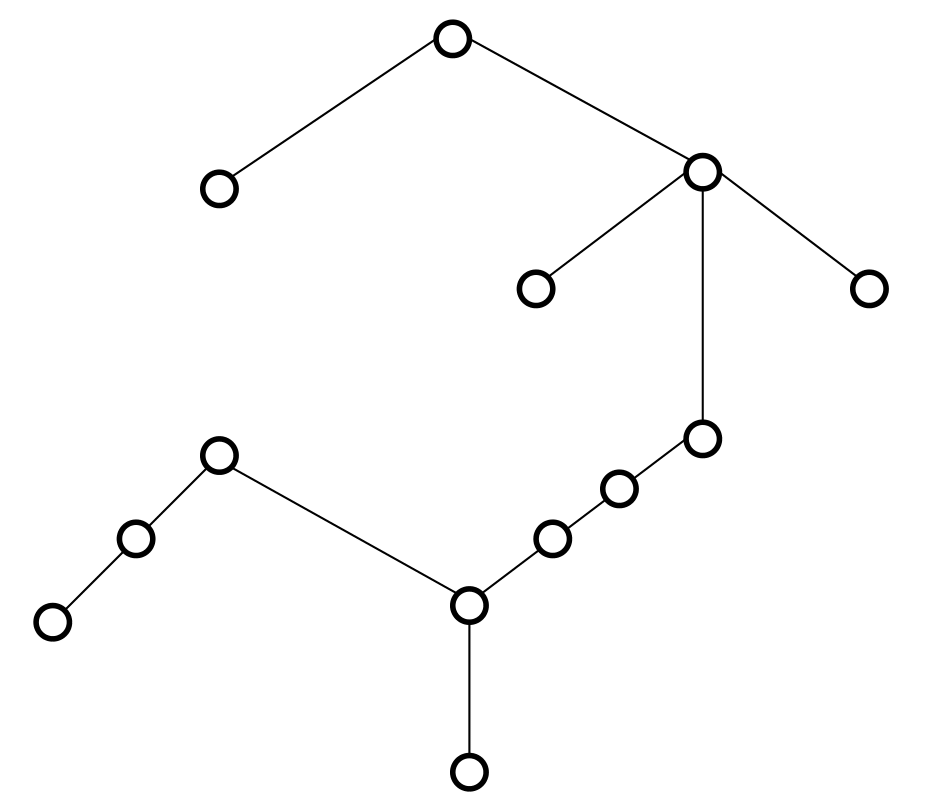}
         \label{subd}     
   }  
        \end{center}
        \caption{(a) A simple graph, (b) a subdivision of the graph given in (a).}
        \label{subdivision}
\end{figure}

\begin{definition}
 A subdivision of a simple graph $G$ is a graph resulting from the subdivision of edges in G.  In other words, a subdivision of a graph is a graph obtained by adding
at least one vertex on an edge of the graph.
\end{definition}
Fig. \ref{subdivision} shows a simple graph (Fig. \ref{simple}) and a particular subdivision  in Fig. \ref{subd}.
  We define a graph induced by the set of the variable nodes of a bipartite graph and then we generalize the definition of subdivision of a graph for bipartite graphs.
\begin{definition}
Let $G(V \cup C,E)$ be a bipartite graph. The simple graph $G'(V,E')$ induced by the set of variable nodes $V$ is a graph with $|V|$ vertices in which two vertices $v_1$ and $v_2$ are connected to each other if and only if there exists a check node $c$ in $C$ such that $v_1$ and $v_2$ are neighbors of $c$.
\end{definition}

\begin{figure}[t]
        \begin{center}
        \subfigure[]{
         \centering\includegraphics[width=1.1in]{ts53cropped1.png}
         \label{1}
         }
        \subfigure[]{
                \centering \includegraphics[width=0.9in]{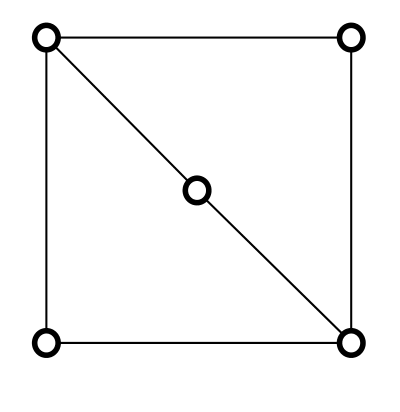}
         \label{2}     
   }  
        \end{center}
        \caption{(a) The (5,3) trapping set as a bipartite graph, (b) the simple graph induced by the 5 variable nodes of the (5,3) trapping set.}
        \label{varinduced}
\end{figure}

As an example, consider the (5,3) trapping set as a bipartite graph. The simple graph induced by the set of variable nodes of the (5,3) trapping set is shown in Fig. \ref{varinduced}.

\begin{definition}
Let ${\cal{T}}(a,b)$ be a trapping set. The trapping set ${\cal{T'}}(a+1,b+1)$ is called a subdivision of  ${\cal{T}}$  if the simple subgraph induced by the set of variable nodes of ${\cal{T'}}$ is a subdivision of  the simple graph induced by the set of variable nodes of ${\cal{T}}$. 
\end{definition}

\begin{figure}[t]
        \begin{center}
        \subfigure[]{
         \centering\includegraphics[width=1.9in]{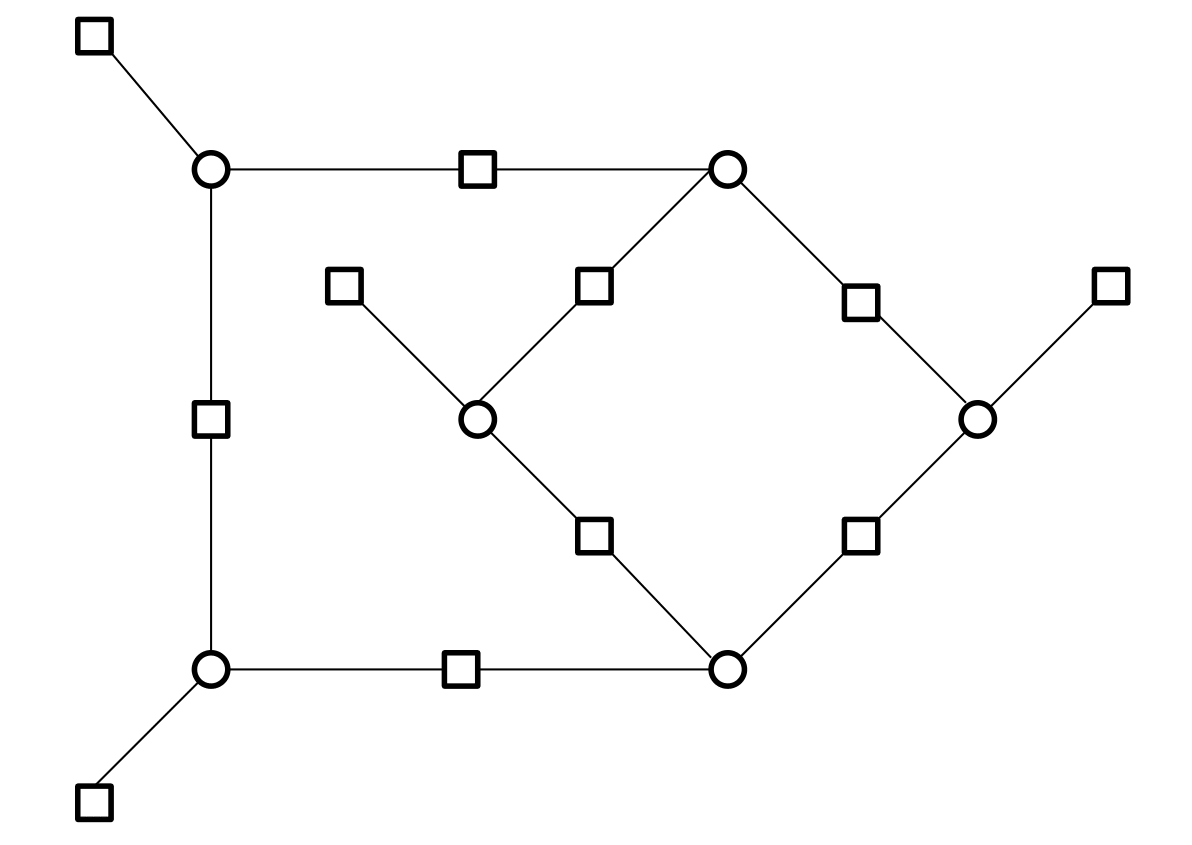}
        
         }
        \subfigure[]{
                \centering \includegraphics[width=1.9in]{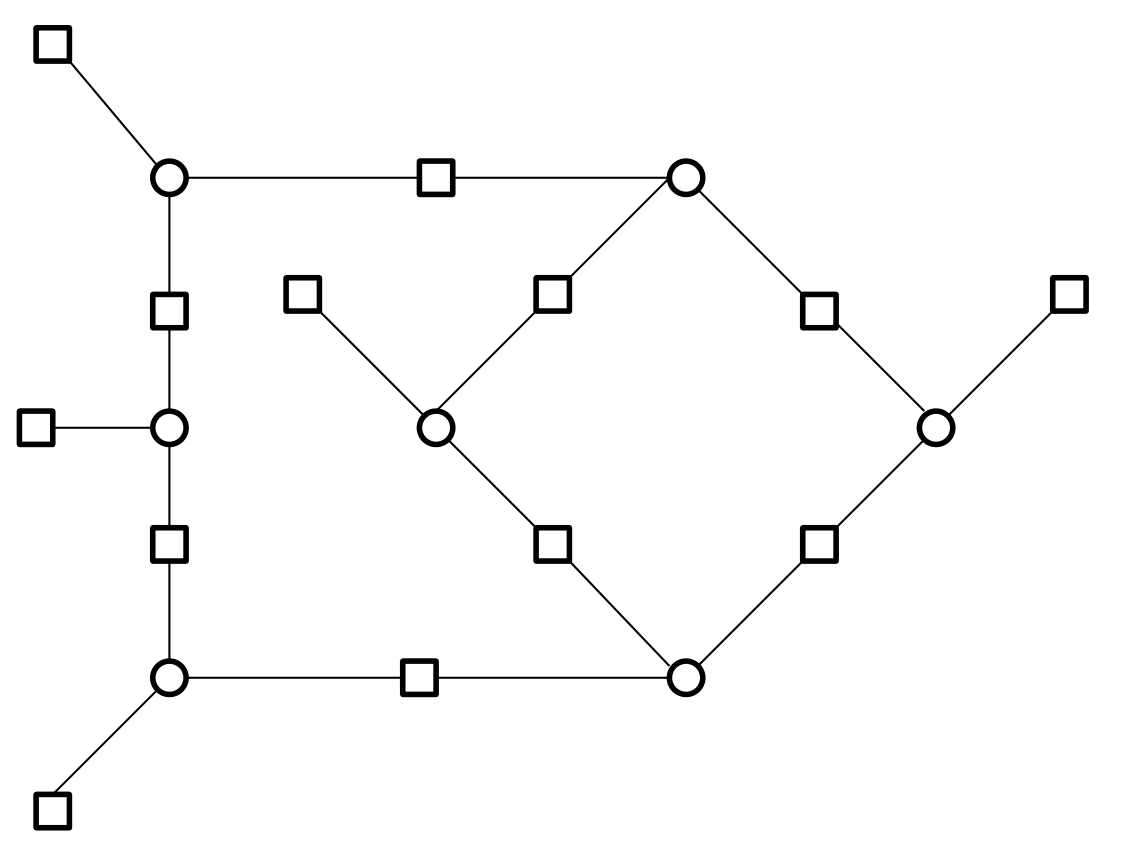}
             
   }  
        \end{center}
        \caption{(a) A (6,4) trapping set, (b) a (7,5) trapping set which is a subdivision of the (6,4) trapping set given in (a).}
        \label{subdivision2}
\end{figure}

Fig. \ref{subdivision2} shows two trapping sets, a (6,4) trapping set and a (7,5) trapping set, in which the (7,5) trapping set is a subdivision of the (6,4) trapping set.

\begin{corollary}
Let ${\cal{T'}}(a+1,b+1)$  be a trapping set which is a subdivision of the trapping set ${\cal{T}}(a,b)$. Then $s_{(a+1,b+1)}({\cal{T'}})=s_{(a,b)}({\cal{T}})$. 
\end{corollary}

As we want to reduce the rate-loss caused by converting single checks to super checks, we now study the minimum number of super checks that are required to be replaced in a Tanner graph of an LDPC code such that the  decoder can correct all error patterns on  all $(a,b)$ trapping sets. 
\begin{definition}
Let ${\cal{C}}$ be a $(3,\rho,8)$-LDPC code with the parity check matrix $H$ and let  ${\cal{T}}(a,b)$ be an elementary trapping set in $H$. 
The minimum number of super checks corresponding to a 2-error correcting component code that are required for  eliminating all $(a,b)$ trapping sets in $H$ is called the {\it{splitting number}} of the $(a,b)$ trapping sets in $H$ and is denoted by $s_{(a,b)}(H)$.
\end{definition}
\subsection{Upper bounds on the splitting number}
Now, we provide  upper bounds on the splitting number of trapping sets in the parity check-matrices based on permutation matrices. Permutation-based LDPC codes  are $(\gamma,\rho)$- regular codes constructed from permutation matrices.  A permutation matrix is any square matrix in which the weight of each row and each column is one. If the permutation matrix is cyclic, the permutation matrix is called a circulant permutation matrix and  the LDPC code becomes quasi-cyclic \cite{Marc}. The parity check matrix of a quasi-cyclic LDPC code can be represented by an array of circulant permutation matrices as follows \cite{Marc}:
\begin{equation} 
\label{matrix}
 H=\left[ \begin{array}{cccc}
I_0 & I_0 & \cdots & I_0 \\
I_0 & I_{p_{1,1}} & \cdots & I_{p_{1,\rho-1}}\\
\vdots & & \ddots & \vdots\\
I_0 & I_{p_{\gamma-1,1}} & \cdots & I_{p_{\gamma-1,\rho-1}} \end{array} \right] 
\end{equation} 
where for $1 \leq j \leq \gamma-1$ and $1 \leq l \leq \rho-1$, $I_{p_{j,l}}$ represents the circulant permutation matrix with a one at column-$(r+p_{j,l})$ mod $p$ for the row $r$ ($0 \leq r \leq p-1$). If for $1 \leq j \leq \gamma-1$ and $1 \leq l \leq \rho-1$, $I_{p_{j,l}}$  is not circulant, then $H$ is just a $(\gamma,\rho)$-regular matrix based on permutation matrices.
\begin{lemma}
\label{Lem1}
Let ${\cal{C}}$ be a $(3,\rho,8)$  LDPC code with the parity-check matrix $H$ based on permutation matrices of size $p$.  Then, $s_{(a,b)}(H)\leq 2p$, for all $a$ and $b$. 
\end{lemma}

Proof: Suppose  the first $2p$ rows  of $H$ are replaced by super checks. The first $2p$ rows of $H$ correspond to the first two rows of blocks in equation (\ref{matrix}). Thus, each variable node is connected to exactly 2 super checks and 1 single parity check. It results that each variable node receives at least 2 correct messages from its neighbors.  In fact, by converting two single parity checks to super checks and then splitting each super check into two single parity check nodes, all cycles in all elementary trapping sets are eliminated. {\it{Q.E.D.}}

According to Lemma \ref{Lem1}, all elementary trapping sets are eliminated when each variable node is connected to exactly two super checks. Thus, the trapping sets for this class of CH-GLDPC codes are non-elementary trapping sets.

We now exhibit a fixed set for the PBF algorithm  for the CH-GLDPC code in the case that the super checks have been replaced such that each variable node is connected to exactly two super checks. 

\begin{theorem}
\label{fixedset}
 Let ${\cal{T}}$ be a subset of variable nodes with the induced subgraph ${\cal{I}}$. Then, ${\cal{T}}$ is a fixed set if (a) The degree of each check node in ${\cal{I}}$ is either 1 or 3 and; (b) Each variable node in ${\cal{I}}$ is connected to 2 check nodes of degree 3 and 1 check node of degree 1 where the check nodes of degree 3 have been replaced by super checks of the 2-error correcting component code and; (c) No 2 check nodes share a variable node outside ${\cal{I}}$. 
\end{theorem}

Proof: Since the check nodes of degree 3 have been replaced by super checks of a 2-error correcting component code and since the decoding in the component codes is the BDD, the super checks of degree 3 do not send any flip messages to the variable nodes in ${\cal{I}}$.  Also, since any variable node in ${\cal{I}}$ is connected to 2 super checks, it remains corrupt. Furthermore, no variable node outside ${\cal{I}}$ receives more than 1 flip message because no 2 check nodes  share a variable node outside ${\cal{I}}$. Thus, the variable nodes outside ${\cal{I}}$ that are originally correct will remain correct. Consequently, ${\cal{I}}$ is a fixed set. {\it{Q.E.D.}}

\begin{figure}[t]
\centering
\includegraphics[width=2.1in]{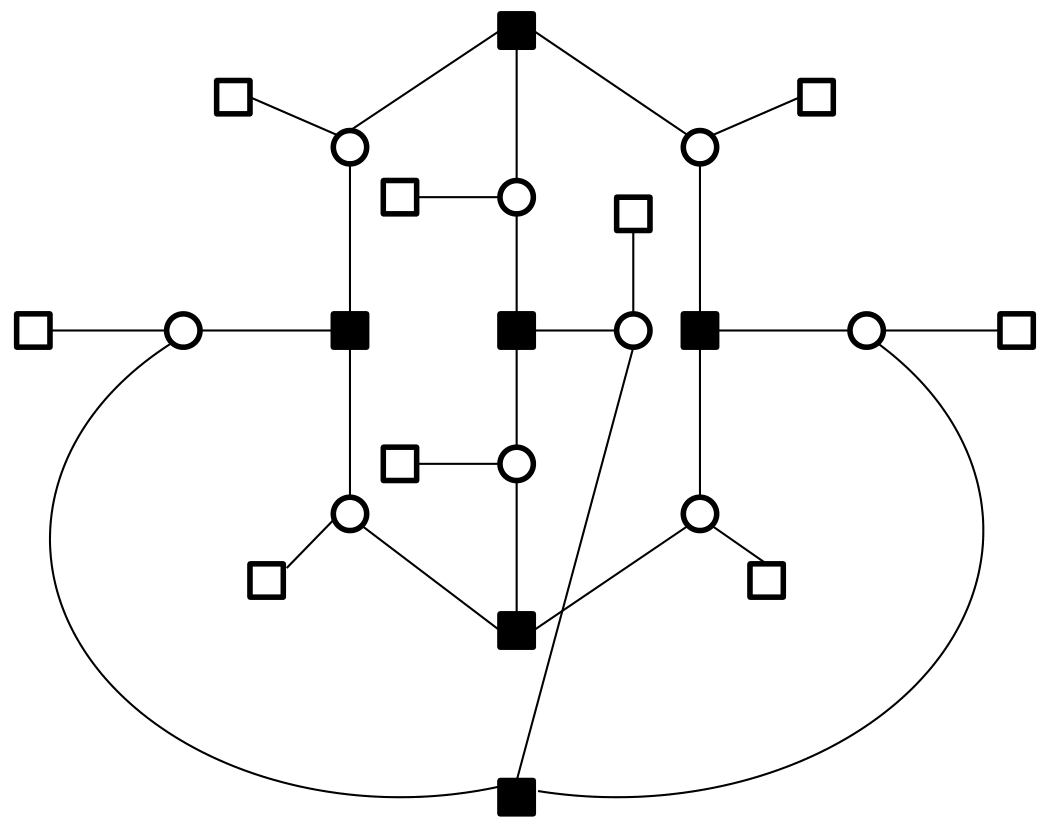}
\caption{A  fixed set for a $(3,\rho,8)$-LDPC code in which each variable node is connected to exactly two super checks.}
\label{fig:fixedset}
\end{figure}

\begin{figure}[t]
\centering
\includegraphics[width=2.1in]{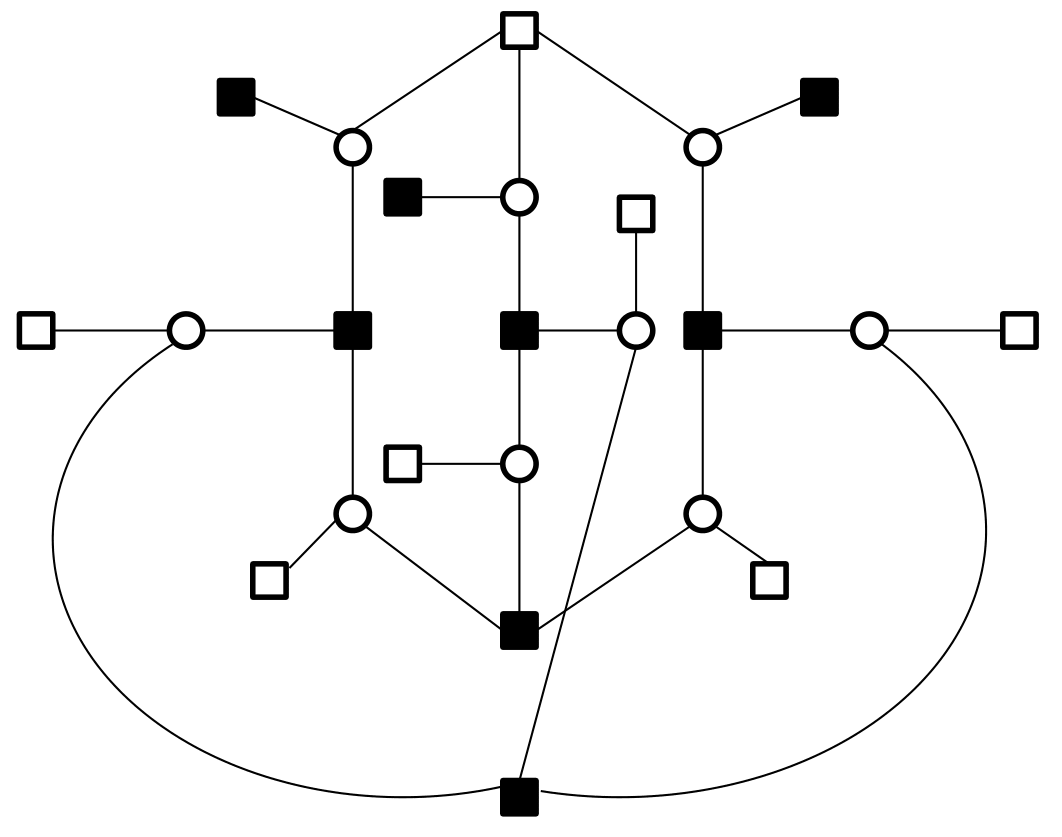}
\caption{An example of a  subgraph in a $(3,\rho,8)$-LDPC code which satisfies  all conditions of Theorem \ref{fixedset} except the condition (b). This structure is not harmful for the PBF algorithm. }
\label{fig:notfixedset}
\end{figure}

Fig. \ref{fig:fixedset} shows a  fixed set in a $(3,\rho,8)$-LDPC code in which each variable node is connected to exactly 2 super checks. We note that conditions (a) and (c) are similar to the corresponding conditions  in Fact 1. The main difference is in condition (b) where in Theorem \ref{fixedset}, the constraint on the position of super checks is a stronger condition on ${\cal{I}}$ to be a fixed set. We also note that if this condition is not satisfied, ${\cal{I}}$ may not be either a trapping set or a fixed set. 
Fig. \ref{fig:notfixedset} shows a subgraph satisfying all conditions of Theorem \ref{fixedset} except the condition (b) which is not a trapping set nor a fixed set. 

\begin{figure}[t]
        \begin{center}
        \subfigure[]{
         \centering\includegraphics[width=1.15in]{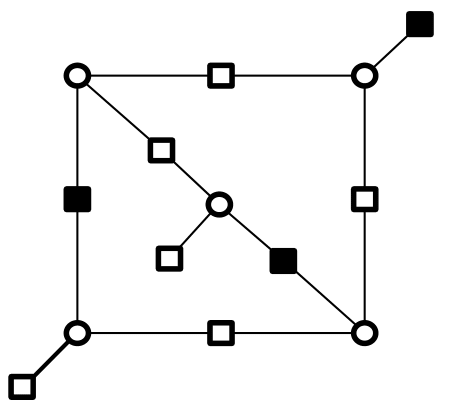}
         \label{ts5,3}
         }
        \subfigure[]{
        \centering \includegraphics[width=1.5in]{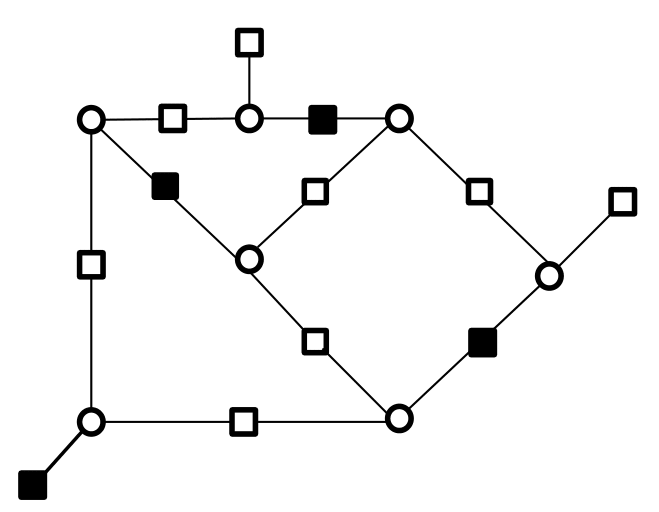}
        \label{ts7,3}
              
   }
         \end{center}
        \caption{Some trapping sets in a column-weight 3 LDPC codes  that can be eliminated if each variable node is connected to exactly one super check. The graphs in (a) and (b)  correspond to the (5,3) and (7,3)  trapping sets, respectively.}
        \label{fig:split}
\end{figure}

Although all elementary trapping sets are eliminated when each variable node is connected to two super checks, there are trapping sets  that are eliminated if each variable node is connected to exactly one super check. Fig. \ref{fig:split} depicts a possible way for replacing super checks in ${\cal{T}}(5,3)$ and  ${\cal{T}}(7,3)$,  such that each variable node is connected to exactly one super check and the trapping sets are not harmful anymore.

Thus, for a permutation-based LDPC code ${\cal{C}}(3,\rho,8)$ with the parity-check matrix $H$, if the parity checks corresponding to the first $p$ rows of $H$ are replaced by super checks, then all  ${\cal{T}}(5,3)$ and ${\cal{T}}(7,3)$  trapping sets are eliminated and hence $s_{(5,3)}(H) \leq p$ and $s_{(7,3)}(H) \leq p$.

\begin{figure}[h]
\centering
\includegraphics[width=1.2in]{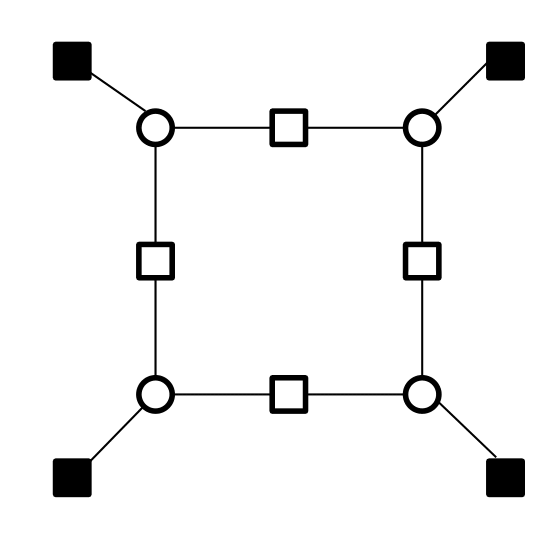}
\caption{The (4,4) trapping set is still harmful if each variable is connected to exactly one super check where have been replaced instead of degree-1 single parity checks in the trapping set.}
\label{ts44harmful}
\end{figure}

 It is easy to see that the smallest trapping set, the (4,4) trapping set, may not be eliminated if each variable node is connected to exactly one super check.  In fact, the $(4,4)$ trapping set will remain harmful if the single parity checks of degree-1 are replaced by super checks (Fig. \ref{ts44harmful}). The following Theorem provides a condition on the parity check matrix $H$ in which all $(4,4)$ trapping sets are eliminated if each variable node is connected to exactly one super check.

\begin{theorem}
\label{Th2}
Let ${\cal{C}}$ be a $(3,\rho,8)$ QC-LDPC code with the parity check matrix $H$. Suppose the first $p$ rows of $H$ are replaced by super checks. Then, $s_{(4,4)}(H) \leq p$ if  the girth of the Tanner graph corresponding to the last $2p$ rows of $H$ is 12.
\end{theorem}
Proof:  If in ${\cal{T}}(4,4)$ the single parity checks of degree-1 are replaced by super checks, then due to the existence of a cycle of length 8, the PBF  cannot correct the errors. However, if the girth of the subgraph induced by the single parity checks is greater than 8, then there will not be any 8-cycle and consequently all  $(4,4)$ trapping sets will be eliminated.  According to Corollary 2.1 in \cite{Marc}, the girth of a $(2,\rho)$-regular QC-LDPC code  is $4i$ for some integer $i>0$. Moreover,  the girth of $H$ cannot be more than 12 as shown in \cite{Marc}. Thus,  if the girth of the subgraph induced by the last $2p$ rows of $H$ is 12, it results that all 8-cycles in $H$ will contain at least one super check of degree 2, and henceforth the 8-cycles are not the harmful $(4,4)$ trapping sets. {\it{Q.E.D.}}

We finish this section by providing a lower bound on the rate of the CH-GLDPC codes.
\begin{lemma}
Let ${\cal{C}}$ be a $(\gamma,\rho)$-regular LDPC code with the parity-check matrix $H_{M \times N}$. Let $C$ be a $t$-error correcting component code of rate $r$ with a full-rank parity-check matrix $H'_{m \times \rho}$. If $\kappa$ be the number of single parity checks in $H$ that are replaced by super checks corresponding to $C$, then the rate of the CH-GLDPC code 
$$R \geq  1-\frac{\gamma}{\rho}-\kappa \lambda (1-r)$$
 where $\lambda = \frac{\rho}{N}$.
\end{lemma}
Proof: If $\kappa$ be the number of super checks that are replaced in $H$, then there will be $(\kappa m+ (M-\kappa))$ rows in the parity check matrix of the CH-GLDPC codes. Thus, the rate of the CH-GLDPC code is:
\begin{eqnarray}
 R &\geq &1- \frac{(\kappa m+ (M-\kappa))}{N} \nonumber\\
  & \geq &1-\frac{\gamma}{\rho}-\kappa (1-r)\frac{\rho}{N}\nonumber
  \end{eqnarray}
  where the last inequality follows from the fact that $1-\frac{M}{N}=1-\frac{\gamma}{\rho}$ and $m-1 <\rho(1-r)$.
  Assuming  $\lambda = \frac{\rho}{N}$ proves the result. {\it{Q.E.D.}} 
  
  \begin{corollary}
	\label{rate}
  Let ${\cal{C}}$ be a $(\gamma,\rho)$-regular LDPC code and let $H_{M \times N}$ be the parity-check matrix based on permutation matrices with size $p$. Let $C$ be a $t$-error correcting component code of rate $r$ with a full-rank parity-check matrix $H'_{m \times \rho}$. If $\kappa=\alpha p$  be the number of single parity checks in $H$ that are replaced by super checks corresponding to $C$, where $\alpha$ is an integer and  $0 \leq \alpha \leq \gamma$, then the rate of the CH-GLDPC code is:
  $$R \geq  1-\frac{\gamma}{\rho}-\alpha (1-r).$$
  \end{corollary}
 
 To see how tight the lower bound on the rate of the CH-GLDPC codes given in Corollary \ref{rate} is consider  a permutation-based ${\cal{C}}(3,31,8)$ LDPC code  of rate 0.9034. If each variable node is connected to 1 super check corresponding to the BCH(31,21), then the actual rate of the CH-GLDPC code is 0.6130 while the lower bound given in Corollary \ref{rate} is 0.5806.
 If each variable node is connected to 2 super checks of the BCH(31,21), then the actual rate is 0.3236 and the lower bound is 0.2580.

 \section{Guaranteed Error Correction Capability of the CH-GLDPC codes } 
\label{GEC}
 In this section, we study the error correction capability of the CH-GLDPC codes in which the global code is a $(3,\rho,8)$ regular LDPC code and the component code is a 2-error correcting code. The code families  that are studied are i) CH-GLDPC codes in which  each variable node is connected to exactly 2 super checks and ii) the CH-GLDPC codes in which each variable node is connected to exactly 1 super check. For simplicity, we denote the first code family with ${\cal{C}}^{I}$ and the second code family with  ${\cal{C}}^{II}$.

\begin{theorem}
\label{GuarErr}
 Consider a CH-GLDPC code \textsc{C} from the code family ${\cal{C}}^{I}$.  Then the PBF can correct up to 5 errors in \textsc{C}.
\end{theorem}
Proof: See Appendix.\\

\begin{figure}[ht]
\centering
\includegraphics[width=2.2in]{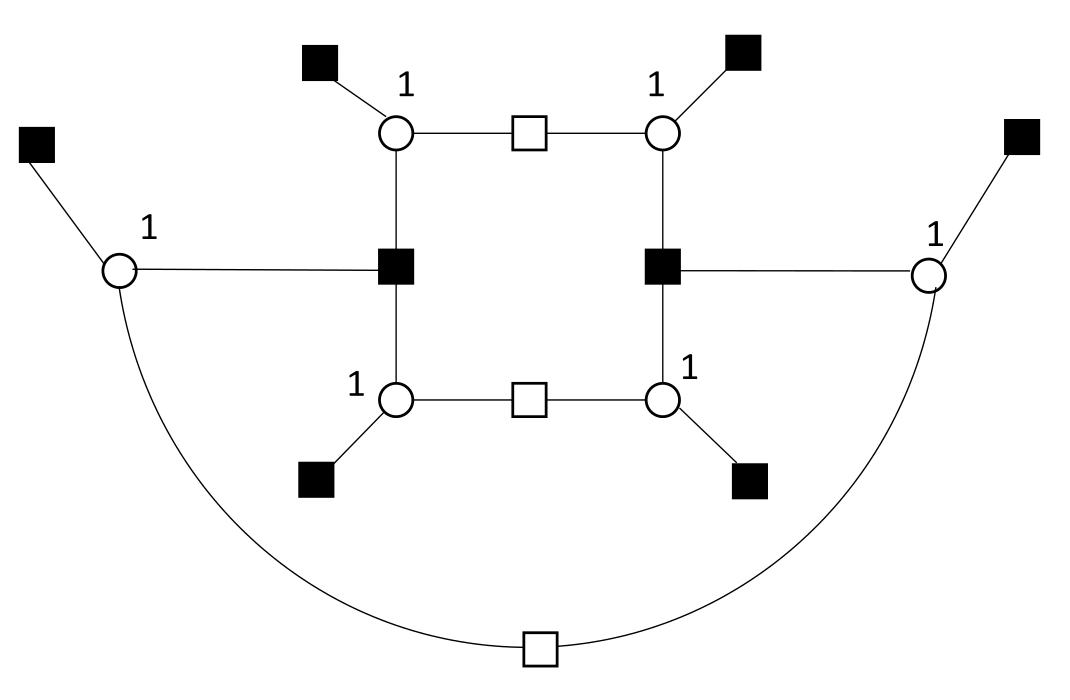}
\caption{An uncorrectable error pattern of size 6 in a CH-GLDPC code that each variable node is connected to 2 super checks.}
\label{6UncorrectableError}
\end{figure}

\begin{corollary}
Consider a CH-GLDPC code \textsc{C} in ${\cal{C}}^{I}$. Then, there exists an error pattern of size 6, in which the PBF fails on correcting the errors.
\end{corollary}
	Proof:	Figure \ref{6UncorrectableError} shows an example in which the PBF fails to correct 6 errors while every variable node is connected to 2 super checks corresponding to
a 2-error correcting component code. {\it{Q.E.D.}}

As shown in Theorem \ref{GuarErr}, when each variable node is connected to 2 super checks of a 2-error correcting component code, then the CH-GLDPC code can correct up to 5 errors. The following Corollary proves the guaranteed error correction capability of the CH-GLDPC codes in ${\cal{C}}^{II}$. 
\begin{corollary}
 Lets suppose  a CH-GLDPC code \textsc{C} in ${\cal{C}}^{II}$. Then the PBF can correct up to 1 error in \textsc{C}.
\end{corollary}
Proof: It is easy to see that if there exist 2 errors on a (4,4) trapping set in which each degree-1 check node is replaced by a super check (as shown in Fig. \ref{ts44harmful}), then the PBF fails. Thus, the guaranteed error correction capability of a CH-GLDPC code in ${\cal{C}}^{II}$  is equal to the error correction capability of PBF for LDPC codes. {\it{Q.E.D.}}

\section{Splitting numbers of ${\cal}(4,\rho,6)$ LDPC codes and trapping sets elimination using  the Gallager B decoding algorithm }
\label{Discussion}
In this section, we generalize our results on critical sets and splitting number of $(3,\rho,8)$ LDPC codes to $(4,\rho,6)$ LDPC codes and  the Gallager B decoding algorithm.
\subsection{Elimination of trapping sets by super checks in $(4,\rho,6)$ LDPC codes}
\begin{figure}[t]
        \begin{center}
       \subfigure[]{
         \centering\includegraphics[width=1in]{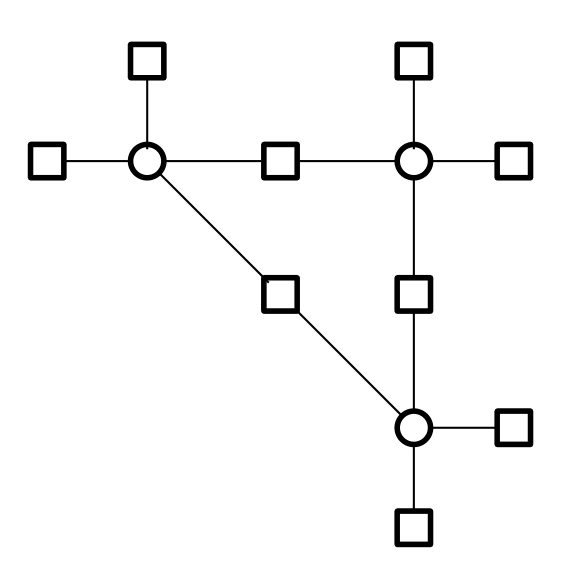}
         \label{ts4,3,6}}
         \subfigure[]{
        \centering \includegraphics[width=0.7in]{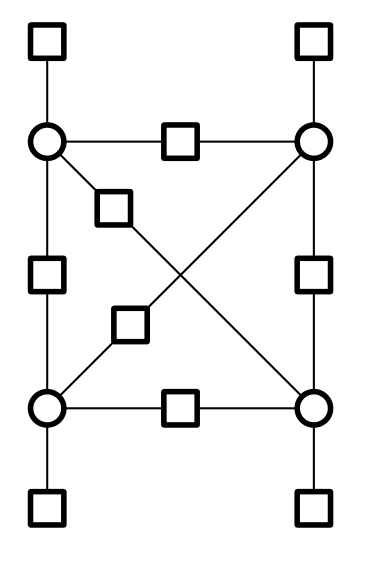}
        \label{ts4,4,4}} 
         \subfigure[]{
         \centering\includegraphics[width=0.97in]{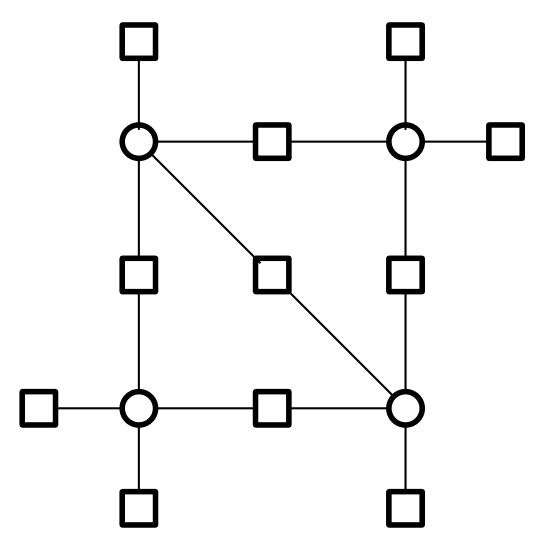}
         \label{ts4,4,6}}
         \end{center}
        \caption{Some small trapping sets in column-weight four LDPC codes with girth 6. (a) the (3,6) trapping set, (b) the (4,4) trapping set and (c) the (4,6) trapping set.}
        \label{TS4}
\end{figure}

\begin{figure}[t]
        \begin{center}
       \subfigure[]{
         \centering\includegraphics[width=1in]{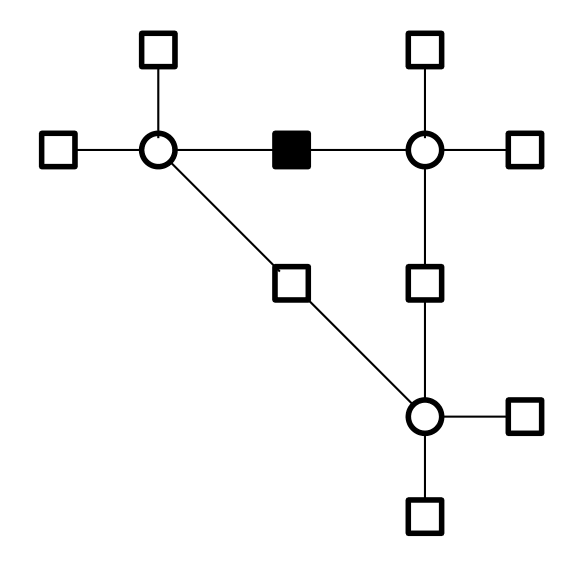}
         \label{ts4,3,6harmless}}
        \subfigure[]{
        \centering \includegraphics[width=0.7in]{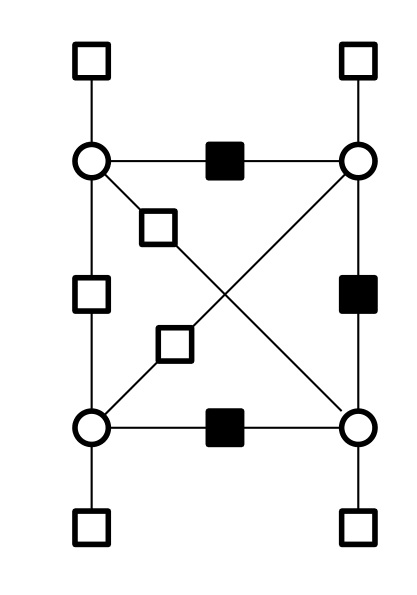}
        \label{ts4,4,4harmless}}
         \subfigure[]{
         \centering\includegraphics[width=0.9in]{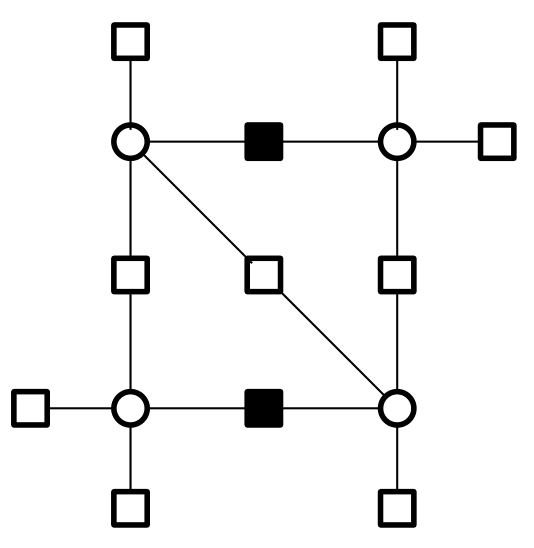}
         \label{ts4,4,6harmless}}
         \end{center}
        \caption{A possible replacement of super checks to eliminate the trapping sets in column-weight four LDPC codes.}
        \label{TS4harmless}
\end{figure}

In Section \ref{SuperChecks}, we provided a method to eliminate harmful (elementary) trapping sets in $(3,\rho,8)$ LDPC codes and provided upper bounds on the splitting number of trapping sets in permutation based LDPC codes. In this section, we extend our results for  $(4,\rho,6)$ LPDC codes.   Fig. \ref{TS4} shows some small trapping sets in a $(4,\rho,6)$ LDPC code.
Fig. \ref{TS4harmless} shows a possible replacement of super checks corresponding to a 2-error correcting component code that eliminates the trapping sets. In Algorithm \ref{Alg3} we provide a  method to find  critical sets in an elementary trapping set of a $(4,\rho,6)$ LDPC code. 

\begin{algorithm}
\caption{Finding a critical set in a trapping set ${\cal{T}}(a,b)$ in a $(4,\rho,6)$ LDPC code.}
\label{Alg3}
\begin{algorithmic}
\STATE {\bf{initialization}:} Let ${\cal{T}}'={\cal{T}}$ be the $(a,b)$ trapping set.
\WHILE{Number of variable nodes in ${\cal{T'}}$ is greater than 0}
\IF{there exists a variable node $v$ in ${\cal{T'}}$ which is connected to two degree-1 check nodes and two degree-2 checks }
\STATE Replace one of the check nodes of degree-2 connected to $v$ by a super check 
  corresponding to a 2-error correcting code. Split the super check into two single checks. Remove the variable node $v$ and all edges connected to it.
\ELSE
\IF{there exists a variable node $v$ in ${\cal{T'}}$ which is connected to one degree-1 check node and three degree-2 checks}
\STATE  Replace two check nodes of degree-2 connected to $v$ by  super checks 
  corresponding to a 2-error correcting code. Split the super checks into two single checks. Remove the variable node $v$ and all edges connected to it.
\ENDIF
\ENDIF
\WHILE{Number of variable nodes connected to at least two single parity checks of degree-1 is greater than 0} 
\STATE Remove variable nodes that are connected to at least two single parity checks of degree-1 and all edges connected to them.
\ENDWHILE
\ENDWHILE
\end{algorithmic}
\end{algorithm}

Following the same methodology used in Section \ref{results}, Lemma \ref{Lem1} can be generalized for column-weight four LDPC codes as follows.

\begin{lemma}
\label{Lem2}
Let ${\cal{C}}$ be a $(4,\rho,6)$  LDPC code with the parity-check matrix $H$ based on permutation matrices of size $p$.  Then, $s_{(a,b)}(H)\leq 3p$, for all $a$ and $b$. 
\end{lemma}

Proof: The proof is similar to the proof of Lemma \ref{Lem1}. If the first $3p$ rows of $H$ are replaced by super check corresponding to a 2-error correcting component code, then each variable node receives at least 3 correct messages from its neighbors and hence all cycles in all trapping sets are broken by the super checks. {\it{Q.E.D.}}

We now present a fixed set for the PBF algorithm  for the CH-GLDPC code in which each variable node in a $(4,\rho,6)$ LDPC code is connected to exactly 3 super checks of a 2-error correcting component code.  
\begin{corollary}
\label{fixedset2}
 Let ${\cal{T}}$ be a subset of variable nodes with the induced subgraph ${\cal{I}}$. Then, ${\cal{T}}$ is a fixed set if (a) The degree of each check node in ${\cal{I}}$ is either 1 or 3 and; (b) Each variable node in ${\cal{I}}$ is connected to 3 check nodes of degree 3 and 1 check node of degree 1 where the check nodes of degree 3 have been replaced by super checks of the 2-error correcting component code and; (c) No 2 check nodes share a variable node outside ${\cal{I}}$. 
\end{corollary}
The following result provides a   condition on the parity check matrix $H$ in which  all $(3,6)$ trapping sets are eliminated if each variable node is connected to one  super check. 
\begin{theorem}
\label{Condition}
Let ${\cal{C}}$ be a $(4,\rho,6)$ QC-LDPC code with the parity check matrix $H$. Suppose the first $p$ rows of $H$ are replaced by super checks. Then, $s_{(3,6)}(H) \leq p$ if  the girth of the Tanner graph corresponding to the last $3p$ rows of $H$ is at least 8.
\end{theorem}

Proof: The proof is similar to the proof of Theorem \ref{Th2}. If each variable node is connected to 1 super check and the girth of the subgraph induced by the single parity checks is greater than 6, then there is not any $(3,6)$ trapping set. {\it{Q.E.D.}}

We may note that under the condition in Theorem \ref{Condition}, the 8-cycles may not broken and so the other small trapping sets shown in Fig. \ref{TS4}, i.e. the $(4,4)$ and the $(4,6)$ trapping sets may  remain harmful. 

\begin{theorem}
\label{GuarErr2}
Let ${\cal{C}}$ be a $(4,\rho,6)$-regular LDPC code. Lets suppose in a CH-GLDPC code constructed using ${\cal{C}}$ as the global code, each variable node is connected to 3 super checks corresponding to a 2-error correcting component code. Then the PBF can correct at least 3 errors in the CH-GLDPC code obtained by replacing super checks.
\end{theorem}
Proof: The proof is similar to the proof of Theorem \ref{GuarErr}. All elementary trapping sets are eliminated when each variable node is connected to 3 super checks. Thus, it is enough to consider the cases that there exists at least one check (single check or super check) connected to more than 2 errors. Recall that since the decoding algorithm of the component codes is the BDD, each super check sends at most two flip messages to the variable nodes in its neighborhood. If a super check is connected to more than 2 corrupt variable nodes, we consider the worst case scenario and assume that the super check sends 2 flip messages to correct variable nodes in its neighborhood. We may note that the errors on a tree subgraph are eventually corrected. It can be easily seen that when a super check or a single check is connected to 3 corrupt variable nodes, all errors are eventually corrected. {\it{Q.E.D.}}

We may note that if in a $(4,\rho,6)$ LDPC code,  each variable node is connected to exactly 1 super check of a 2-error correcting component code, or if is connected to exactly 2 super checks, the error correction capability of the CH-GLPDC code is equal the error correction capability of a $(4,\rho,6)$ LDPC code which is 2. This is due to the fact that in these classes of CH-GLDPC codes, the smallest trapping sets in a  $(4,\rho,6)$ LDPC code are not necessarily eliminated.

\subsection{Elimination of trapping sets by super checks using Gallager B decoding algorithm}
In this section, we show that the method used for eliminating the trapping sets of a column-weight 3 LDPC code with the PBF algorithm can also be used for eliminating the trapping sets with the Gallager B decoding algorithm. To show how the results of the PBF algorithm can be generalized for the Gallager B decoding algorithm, we first explain the  decoding algorithm of the CH-GLDPC codes using the Gallager B decoding algorithm on the global code and the BDD on the component codes in Algorithm \ref{Alg4}.
 \begin{algorithm}
\caption{The Gallager B decoding algorithm for CH-GLDPC codes.}
\label{Alg4}
\begin{algorithmic}
\STATE {\bf{Initialization}}
The variable nodes send their received values to the
neighboring single checks and super checks over the incident edges.
\STATE {\bf{In each iteration:}}
\STATE {\bf{~~Updating rule at check nodes:}}
 \begin{itemize}
 \item Each super check node performs the BDD on the incoming messages. If a codeword is found, then the check node sends  the values of the BDD decoder to the variable nodes. If not, then the check node  sends the value of each variable node to itself.
 \end{itemize}
 \begin{itemize}
 \item At each single parity check, the message sent from a check to a neighboring variable is the sum of all incoming messages except the one arriving from the variable. 
 \end{itemize}
\STATE {\bf{~~Updating rule at variable nodes:}}
\begin{itemize} \item The
message sent from a variable to a neighboring check is the
majority (if it exists) among all incoming messages except the
one arriving from the check. If a majority does not exist, then
the received value corresponding to the variable is sent to the
check.
\end{itemize}
\end{algorithmic}
\end{algorithm} 

\begin{figure}[t]
        \begin{center}
       \subfigure[]{
         \centering\includegraphics[width=1in]{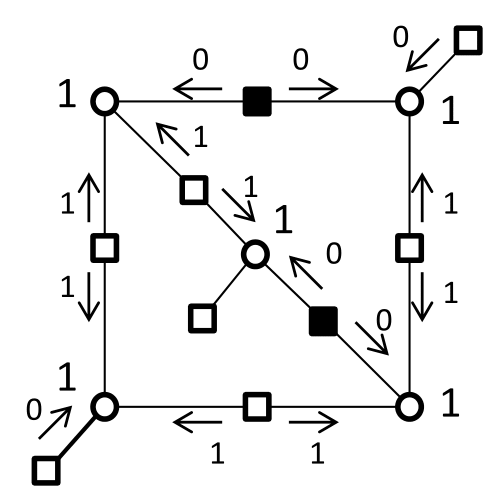}
         \label{GalBts5,3}}
         \subfigure[]{
        \centering \includegraphics[width=1in]{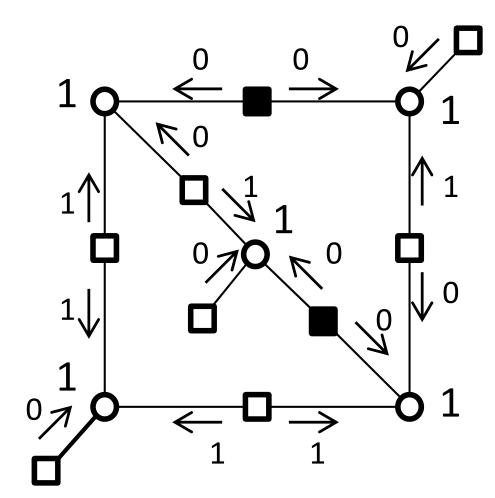}
        \label{GalBts5,32}} 
         \subfigure[]{
         \centering\includegraphics[width=0.97in]{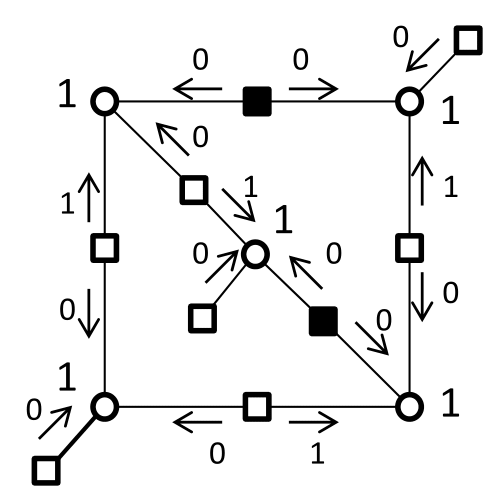}
         \label{GalBts5,33}}
         \subfigure[]{
        \centering \includegraphics[width=1in]{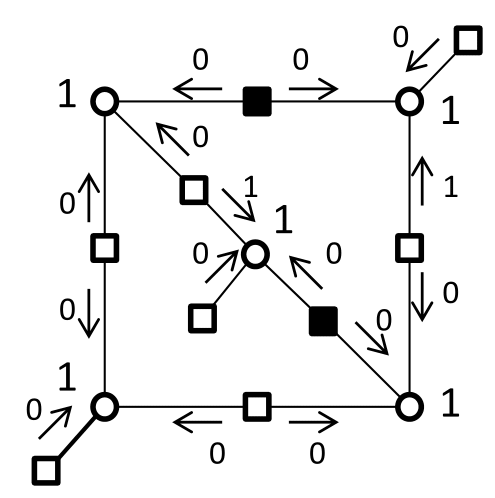}
        \label{GalBts5,34}} 
         \subfigure[]{
         \centering\includegraphics[width=0.97in]{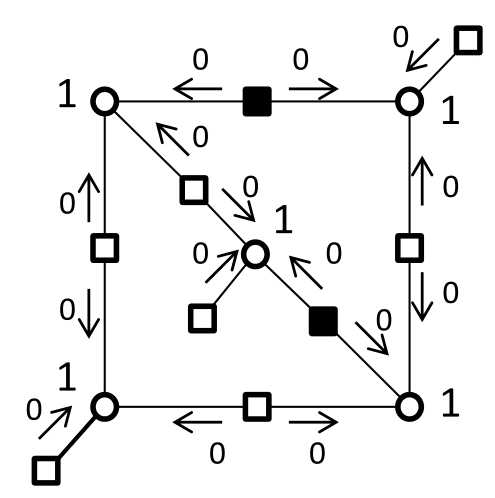}
         \label{GalBts5,35}}
          \subfigure[]{
         \centering\includegraphics[width=0.97in]{ts534cropped1.png}
         \label{GalBts5,36}}
         \end{center}
        \caption{The (5,3) trapping set in a column-weight three code that is eliminated if two super checks corresponding to a 2-error correcting component
code are replaced. Arrows show  messages from check nodes to 
variable nodes in each iteration of the Gallager B decoding algorithm: (a)-(d)  messages from
checks in  iterations 1 to 4, (e) all variable nodes are corrected after the fourth iteration.}
        \label{GalB}
\end{figure}

Fig. \ref{GalB} shows how the Gallager B decoding algorithm can correct all errors on the $(5,3)$ trapping set.

 It is easy to see that the role of a super check of a 2-error correcting component code in  trapping set using Gallaber B decoding algorithm  is similar to the role of a super check of a 2-error correcting component code in  trapping set using the PBF. Hence, by carefully replacing the super checks in the trapping set, the cycles responsible for the failure of the Gallager B decoder are broken. Thus, the results obtained for finding a critical set and the upper bounds on the splitting number of the trapping sets  with the PBF are also correct for the Gallager B decoder. For the guaranteed error correction capability of the CH-GLDPC codes using the Gallager B decoding, it can be easily seen that the  Theorem \ref{GuarErr} is also true for the Gallager B decoding algorithm. A single check that sends a flip message to variable nodes in the PBF, sends 0 to a variable node that is in error and sends 1 to a correct variable node in the Gallager B decoding algorithm. Thus, the same analysis used in the proof of Theorem \ref{GuarErr} can be used to prove it for the Gallager B decoding algorithm.

 The subgraph  shown in Fig. \ref{6UncorrectableError} is also an  error pattern of size 6 for the failure of the Gallager B decoding algorithm. We note that in the CH-GLDPC codes in which each variable node is connected to exactly 1 super check, the error correction capability of the code with the Gallager B decoding algorithm is the same as the error correction capability of a $(3,\rho,8)$ LDPC code with the Gallager B decoding. In this case, the error correction capability of the CH-GLDPC code is 2.


\section{Conclusion }
\label{conclusion}
In this paper, we introduced a method for constructing CH-GLDPC codes in which the super checks corresponding to a 2-error correcting component code are chosen based on the knowledge of trapping sets of a column-weight 3 global LDPC code. By replacing the super checks, we eliminated harmful trapping sets of the PBF algorithm while minimizing the rate loss caused by adding more constraints on check nodes of the component code. We also studied the guaranteed error correction capability in the CH-GLDPC codes. The results were also extended to the Gallager B decoding algorithm and column-weight 4 LDPC codes. 


\begin{appendix}
Proof of Theorem \ref{GuarErr}: To prove this theorem, we first note that according to Lemma \ref{Lem1}, all elementary trapping sets are eliminated when each variable node is connected to two super checks. Thus, it is enough to consider the cases that there exists at least one check (single check or super check) connected to more than 2 errors. We also mention that since the decoding algorithm of the component codes is the BDD, each super check sends at most two flip messages to the variable nodes in its neighborhood. If a super check is connected to more than 2 corrupt variable nodes, we consider the worst case scenario and assume that the super check sends 2 flip messages to correct variable nodes in its neighborhood. We may note that the errors on a tree subgraph are eventually corrected. In other words, every trapping set must contain at least one cycle. 
Recall that as we showed in Fig. \ref{cycle breaking}, super checks corresponding to a 2-error correcting component codes break the cycle if they are connected to at most 2 corrupt variable nodes. Using these facts, we show that all error patterns of size 5 are corrected as their corresponding subgraph can be transformed to a tree. 

We first consider all  possible subgraphs in which a super check node is connected to more than 2 corrupt variable nodes and show that the subgraphs with different error patterns can be transformed to a tree. Then, we repeat it for a single check connected to more than 2 corrupt variable nodes.  To construct all the subgraphs in which a super check node is connected to more than 2 corrupt variable nodes, we consider a super check as a root (level 1) and expand it. The root check node is connected to at least 3 corrupt variable nodes to this super check (level 2). As we mentioned before, we consider the worst case scenario and assume that the super check node sends 2 flip messages to 2 correct variable nodes. The other variable nodes connected to the root check node always send 0 to the root check node, therefore, it is sufficient to connect the root check node to the corrupt variable nodes and two correct variable node to which the flip massages are sent. We then expand this graph by connecting 2 check nodes (one single and one super check) to each variable node of level 2 to construct the level 3 check nodes.  We note that since girth of the global code is 8, a cycle can only be made in at least 5th level of constructing the subgraph. However, as we show, the cycles are broken due to super check nodes connecting to at most two corrupt variable nodes and the graph forms a tree.

\begin{figure}[t]
        \centering\includegraphics[width=1.8in]{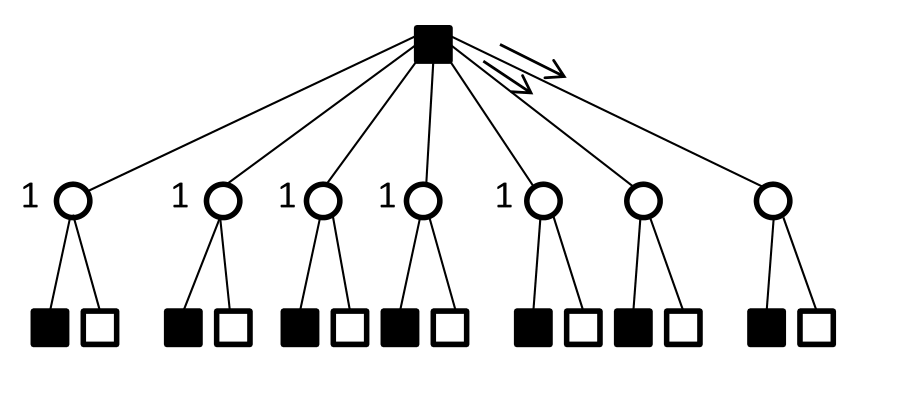}
			\caption{A correctable error pattern of size 5 in which all corrupt variable nodes are connected to one super check.}
         \label{5Err_5}
\end{figure}

\begin{figure}[t]
        \centering\includegraphics[width=1.8in]{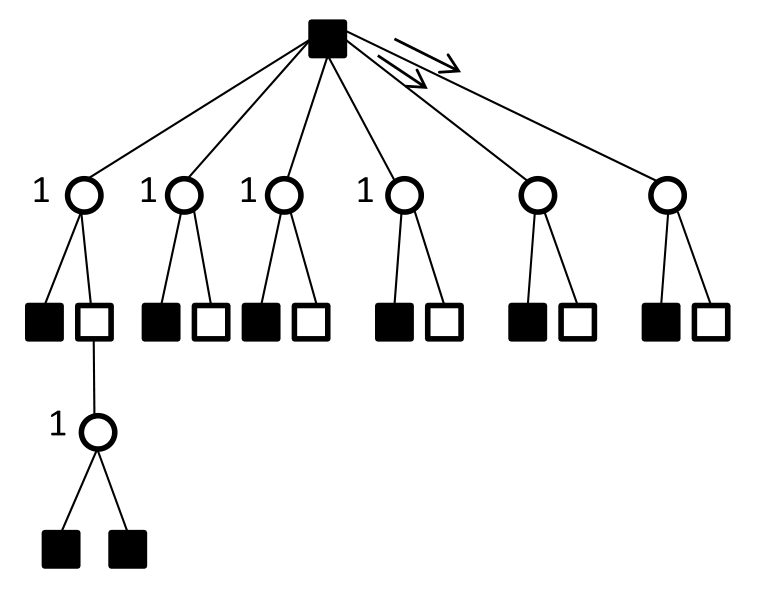}
				\caption{A correctable error pattern of size 5 in which 4 corrupt variable nodes are connected to one super check and one error is connected to one single check.}
        \label{5Err_4}
\end{figure}

Figure \ref{5Err_5} shows how 5 corrupt variable nodes that are connected to one super check are corrected in one iteration.
 Fig. \ref{5Err_4} is one of the cases that an error pattern of size 5 is considered in which 4 errors are connected to one super check and  one error is connected to one single check.
 It is easy to see that all the other error patterns in which 4 errors are connected to one super check can be corrected in at most 2 iterations since this graph is transformed a tree.

\begin{figure}[ht]
\begin{center}
        \subfigure[]{
       \centering \includegraphics[width=2.6in]{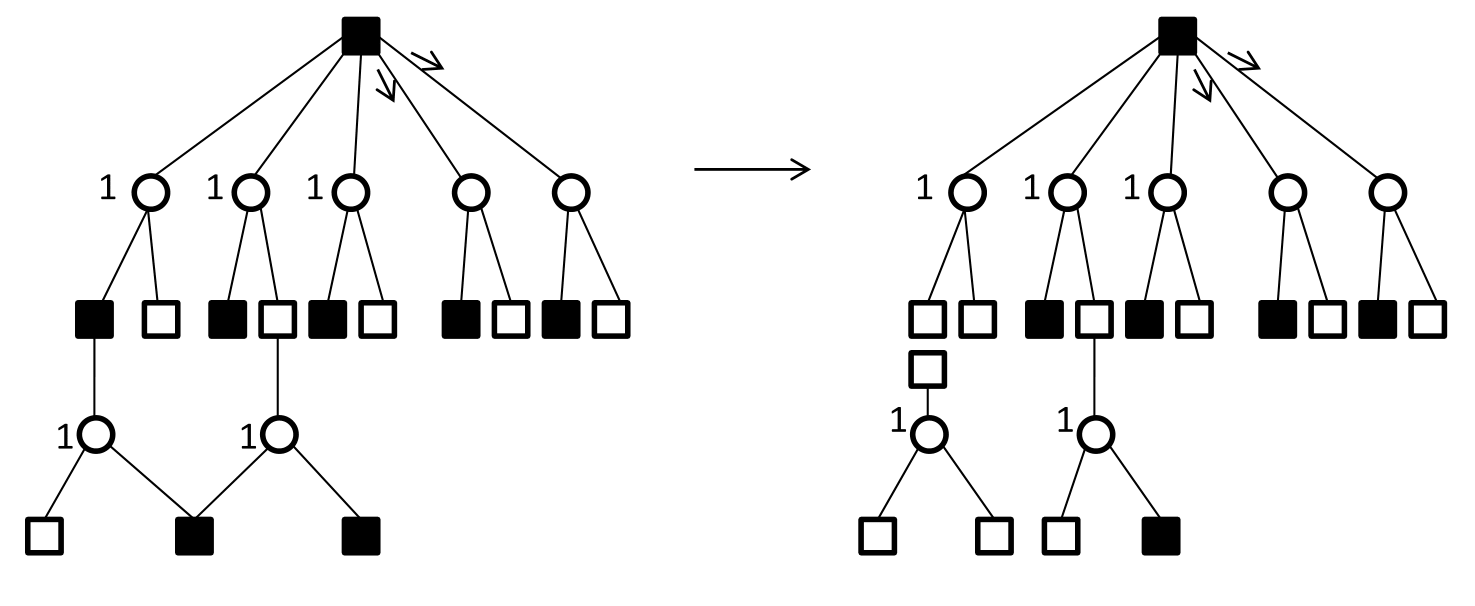}
     \label{5Err_3_1}} 
        \subfigure[]{
        \centering\includegraphics[width=2.6in]{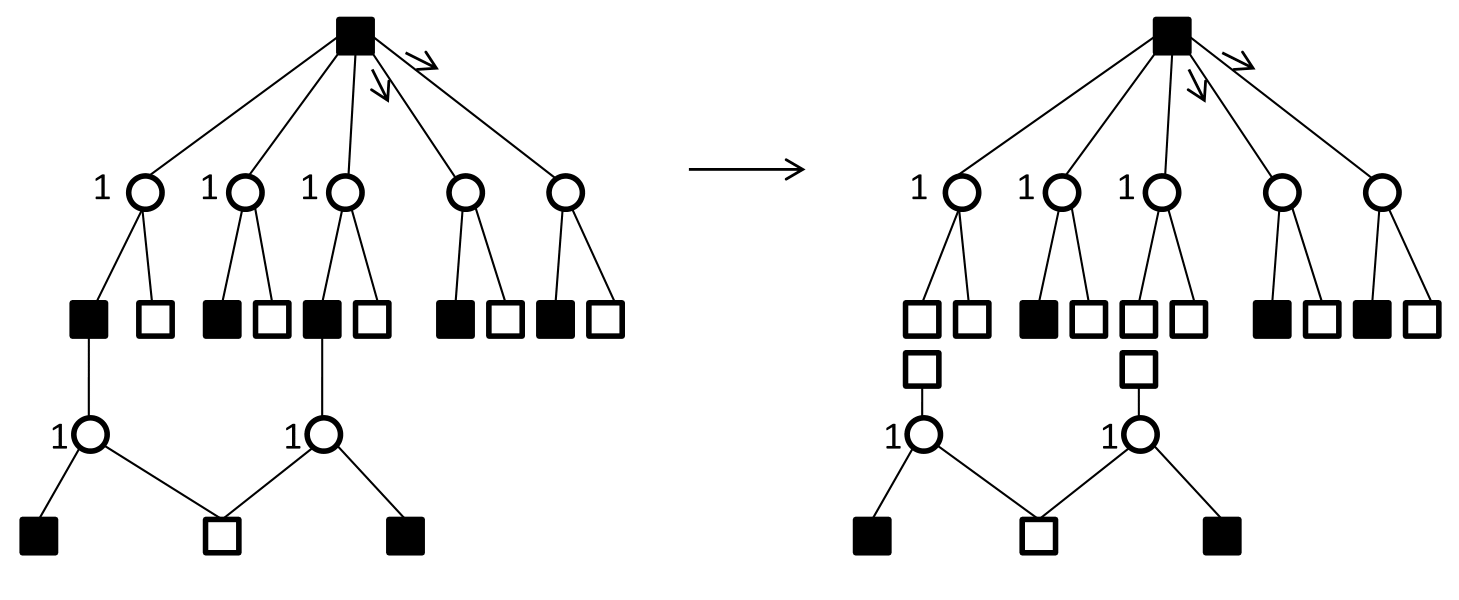}
        \label{5Err_3_2}}
        \subfigure[]{
       \centering \includegraphics[width=2.6in]{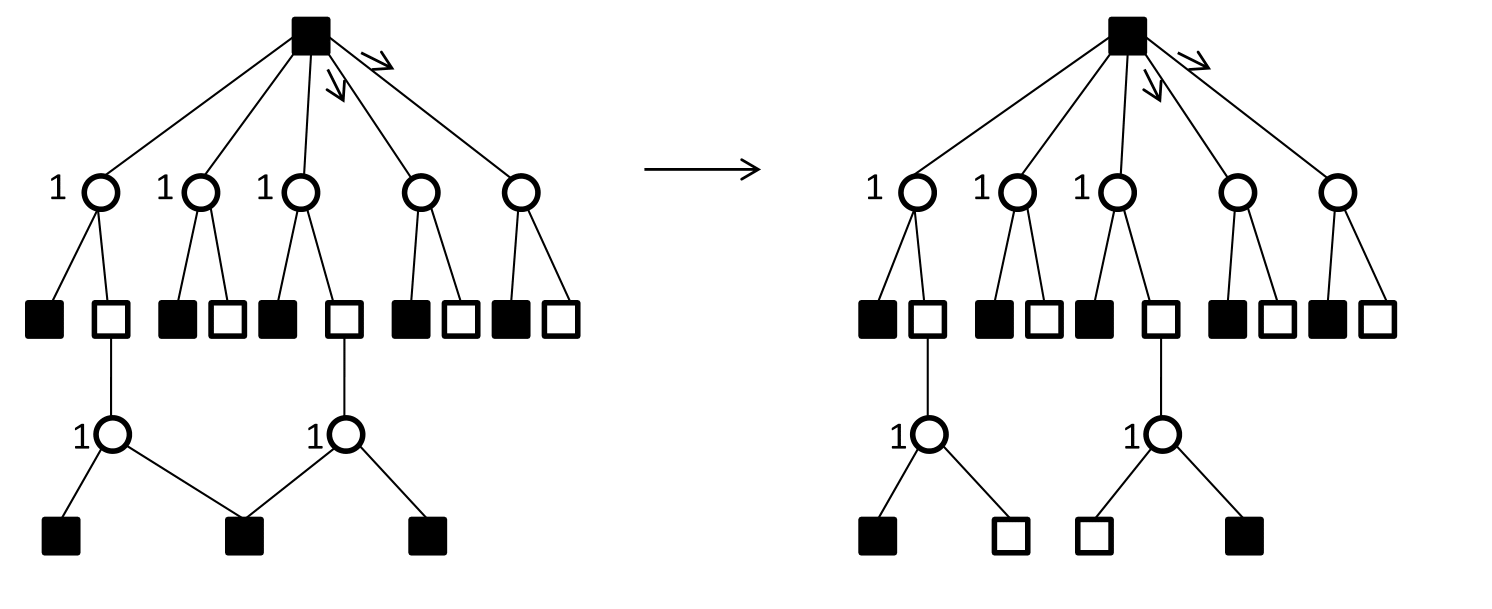}
     \label{5Err_3_3}} 
     \subfigure[]{
       \centering \includegraphics[width=2.6in]{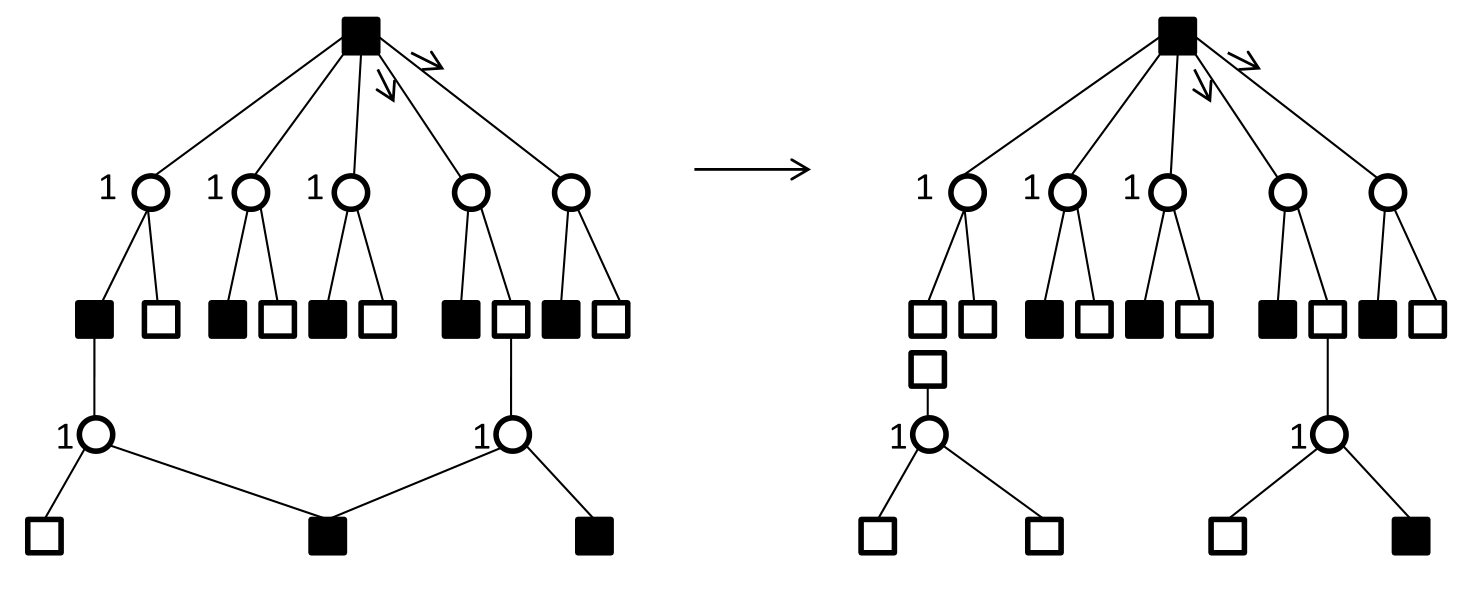}
     \label{5Err_3_4}} 
     \subfigure[]{
       \centering \includegraphics[width=2.6in]{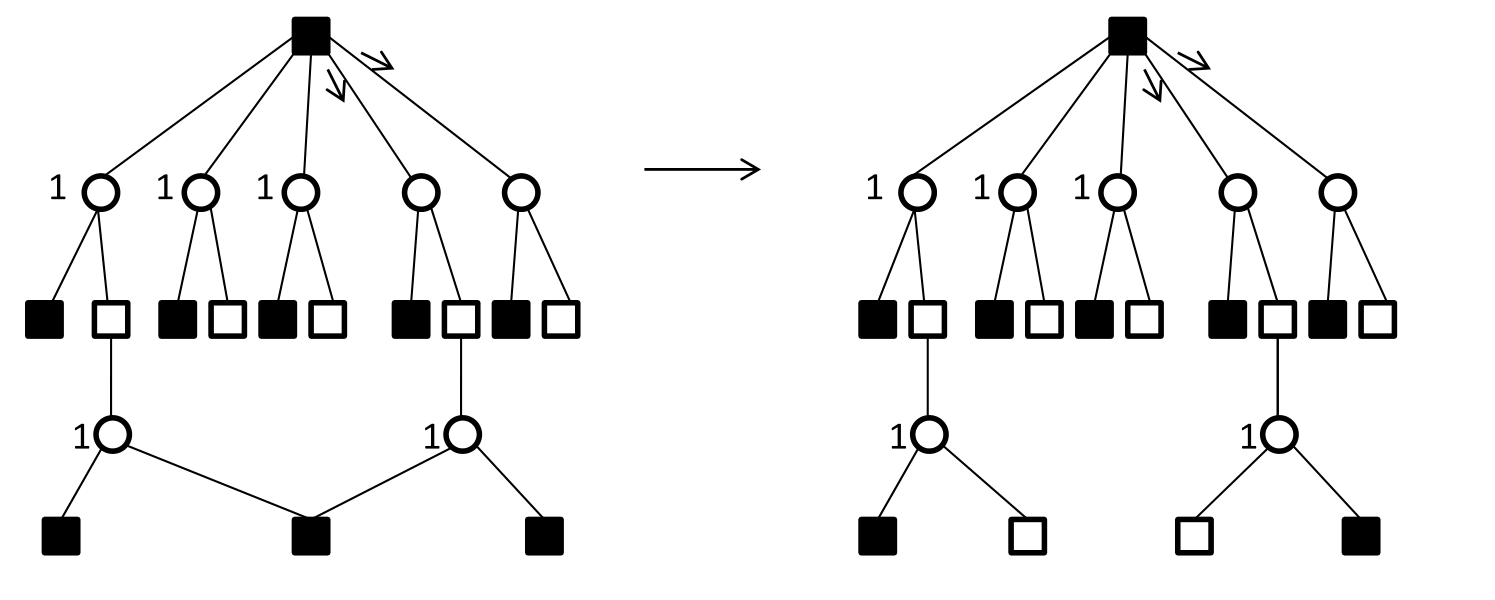}
     \label{5Err_3_5}}
     \subfigure[]{
       \centering \includegraphics[width=2.6in]{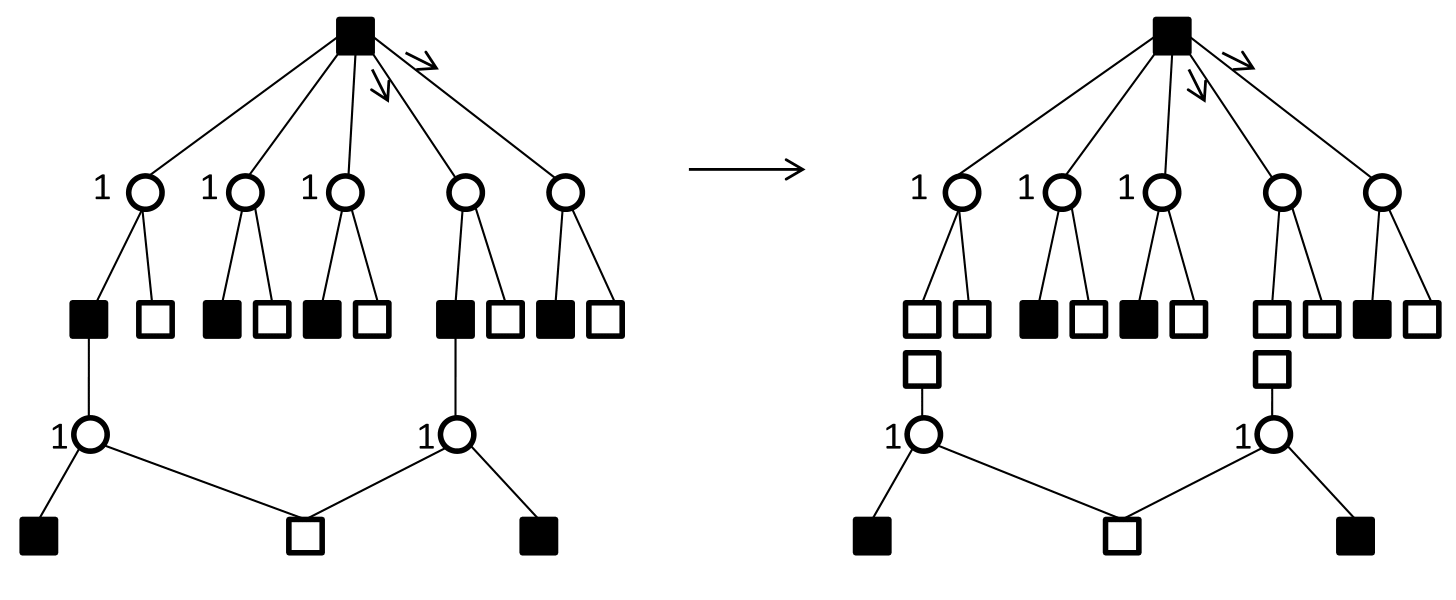}
     \label{5Err_3_6}}  
        \caption{Some possible correctable error patterns of size 5 in which one super check is connected to 3 corrupt variable nodes. }
       \label{3errors}
       \end{center}
 \end{figure}

In Fig. \ref{5Err_3_1}, 	an error pattern of size 5 is considered in which 3 errors are connected to one super check and  2 errors are connected to one single check and one super check. The cycle is broken when each super check is replaced by 2 degree-1 single check.  Figures \ref{5Err_3_2}-\ref{5Err_3_6}	show other possible error patterns of size 5 in which 3 errors are connected to one super check. 	

Now, we consider the case that one single check is connected to more than 2 corrupt variable nodes and all the other super check nodes are connected to at most two corrupt variables. Therefore, all the super checks can break the cycles and the errors are corrected in one iteration. Note that the case in which one single check and one super check are connected to more than 2 corrupt variable nodes, is already included in the previous cases (super check as the root check node).
\end{appendix}


\nopagebreak
\end{document}